\documentclass[11pt]{article}

\usepackage{capt-of}
\usepackage[]{algorithm2e}
\usepackage{amssymb}
\usepackage{amsfonts}
\usepackage{amsmath,amsthm}
\usepackage[nohead]{geometry}
\usepackage[singlespacing]{setspace}
\usepackage{indentfirst}
\usepackage{bbm}
\usepackage{graphicx}
\usepackage{subcaption}
\usepackage{float}
\usepackage{authblk}
\usepackage{multirow}
\usepackage[english]{babel}
\usepackage[usenames, dvipsnames]{color}
\usepackage[strings]{underscore}
\usepackage{url}

\newcommand{\cA}{\mathcal A}
\newcommand{\cC}{\mathcal C}
\newcommand{\cM}{\mathcal M}
\newcommand{\cL}{\mathcal L}
\newcommand{\CD}{\mathcal D}
\newcommand{\cH}{\mathcal H}
\newcommand{\cF}{\mathcal F}
\newcommand{\bI}{\mathbf I}
\newcommand{\bP}{\mathbf P}
\newcommand{\cS}{\mathcal S}
\newcommand{\cP}{\mathcal P}
\newcommand{\cG}{\mathcal G}
\newcommand{\E}{\mathbb E}

\newcommand{\bx}{\mathbf{x}}
\newcommand{\by}{\mathbf{y}}

\usepackage{accents}
\newlength{\dhatheight}

\newcommand{\ti}{{t_k}}
\newcommand{\tii}{{t_{k+1}}}
\newcommand{\ts}{{t_s}}
\newcommand{\tss}{{t_{s+1}}}

\DeclareMathOperator\argmin{arg\, min}
\DeclareMathOperator\argmax{arg\, max}
\DeclareMathOperator\argsup{arg\, sup}
\DeclareMathOperator\spn{span}

\theoremstyle{remark}
\newtheorem{remark}{Remark}

\begin{document}

\title{Simulation Methods for Stochastic Storage Problems: A Statistical Learning Perspective}
\date{\today}
\author{Michael Ludkovski\thanks{Email: ludkovski@pstat.ucsb.edu} \ and Aditya Maheshwari\thanks{Email: maditya0310@gmail.com. \\ This work is partially supported by NSF DMS-1736439.  } }

\affil{Department of Statistics and Applied Probability, South Hall, University of California, Santa Barbara, CA 93106 USA.}

\maketitle

\begin{abstract}
We consider solution of stochastic storage problems through regression Monte Carlo (RMC) methods. Taking a statistical learning perspective, we develop the dynamic emulation algorithm (DEA) that unifies the different existing approaches in a single modular template. We then investigate the two central aspects of regression architecture and experimental design that constitute DEA. For the regression piece, we discuss various non-parametric approaches, in particular introducing the use of Gaussian process regression in the context of stochastic storage. For simulation design, we compare the performance of traditional design (grid discretization), against space-filling, and several adaptive alternatives. The overall DEA template is illustrated with multiple examples drawing from natural gas storage valuation and optimal control of back-up generator in a microgrid. \end{abstract}


\section{Introduction}

Stochastic storage problems concern the optimal use of a limited inventory capacity under uncertainty, motivated by models from commodity management, energy supply, operations research and supply chains. The common thread in all these settings is deciding how to optimally add and reduce inventory as the system state stochastically fluctuates over time. For example, in the gas storage version \cite{bauerle16, boogert08,boogert12, ludkovski10, forsyth07, malyscheff17, boogert13, rasmussen09,warin17,warin}, the objective is to manage an underground cavern through buying and selling natural gas, with the principal stochastic factor being the commodity price. In the microgrid context \cite{heymann17,heymann16,heymann16_aging}, the objective is to deliver electricity at the lowest cost by maximizing inter-temporal storage linked to a renewable intermittent power source (say from solar or wind), so as to minimize the use of non-renewable backup generator. In the hydropower pumped storage setup, the objective is to match upstream inflows and downstream energy demands at the lowest cost \cite{delara17,davison12,davison09}.

To study a stochastic storage problem, one postulates the dynamics of the stochastic risk factors and defines the objective functional. The solution is then obtained by Dynamic Programming, reducing the global optimization into an iterative sequence of local problems, solved via backward recursion in time. There are two main approaches used in the literature to handle these control problems: probabilistic simulation based methods \cite{bauerle16, boogert08,boogert12, ludkovski10,denault2013, felix12, boogert13, warin} and partial differential equations (PDE) \cite{forsyth07,thompson16}.

In this study, we focus on simulation strategies, specifically the regression Monte Carlo (RMC) framework. Introduced for American option pricing \cite{ls2001,tvr}, RMC was adapted to storage problems in \cite{boogert08,ludkovski10,denault2013,cong16,balata17,nadarajah17}. The main complication is the presence of the endogenous inventory variable whose trajectory depends on the control. Thus, storage problems are categorized into a more general class known as optimal switching, where the controller of the facility can switch between different regimes \cite{Pham2009}. Since the storage problem is solved backward in time and the controller does not know which states will be reached, the original path-based Longstaff-Schwartz strategy in \cite{ls2001} is not feasible. To overcome this hurdle, Carmona and Ludkovski \cite{ludkovski10} suggested to back-propagate the inventory trajectories backward in time. This direction was further explored in \cite{balata17,denault2013}. A different approach is to view the overall storage problem as a continuum collection of switching problems, indexed by the current level of inventory $I_t$. One may then reduce to a finite number of 1D switching problems by discretizing the inventory $I$, which are then solved in parallel, employing interpolation in the $I$-direction  \cite{boogert08,boogert13}.

An alternative is to view the stochastic risk factor $P$ and the inventory $I$ jointly in a multivariate setup, either globally \cite{balata17, boogert13}, or locally \cite{warin12, warin} by partitioning the $(P,I)$ domain into sub-domains. This was further extended in \cite{langrene14,langrene15} who also incorporated the control $c$ into the state to handle problems where inventory dynamics feature further exogenous shocks. To deal with the unknown state distribution, the authors suggested multiple iterations of the backward procedure to improve the quality of the solution.

Beyond RMC, other Monte Carlo-based alternatives include the recent proposal in Van-Ackooij and Warin \cite{warin17} to apply Stochastic Dual Dynamic Programming and the Markov Decision Process approach using binomial or multinomial trees \cite{bauerle16,felix12}.

The RMC valuation algorithms  can be divided into three sub-problems:
\begin{itemize}
    \itemsep-0.3em
    \item Approximation of the conditional expectation/continuation value;
    \item Evaluation of the optimal control.
    \item Estimation of the pathwise value.
\end{itemize}
In the storage setup most models consider discrete storage regimes, so that the optimal control is simply the maximizer (obtained generally via brute force comparison) of the continuation value. As for the conditional expectation, within the umbrella of RMC, there are three main methods popular in the literature to approximate it: Regress Now Monte Carlo (RNMC) \cite{boogert08,boogert12,ludkovski10, ls2001, boogert13,  warin, tvr}, Regress Later Monte Carlo (RLMC) \cite{balata17,cong16,jain12,jain15,nadarajah17} and Control Randomization (CR) \cite{langrene14,langrene15}. Computation of the pathwise values can be based on 1-step look-ahead (the so-called TvR scheme introduced in \cite{tvr}), $w$-step look-ahead utilizing the partial trajectories as proposed by Egloff in a series of papers \cite{egloff05,egloff07} or a full $[t,T]$-trajectory as introduced by Longstaff and Schwartz \cite{ls2001}.

\subsection*{Contributions}
In this article we present a unified treatment of stochastic storage problems from the statistical learning perspective.  We recast simulation-based  dynamic programming approaches as an iterative sequence of machine learning tasks. At its basic level, the tasks correspond to approximating the value functions, indexed by the time-step parameter $t$. Equivalently, the tasks can be thought of as learning the underlying $q$-values, or the optimal feedback control $c_t$, both useful alternative views.

With the machine learning perspective, the storage model is viewed as a stochastic simulator that produces noisy pathwise observations, and the aim is to recover the latent \emph{average} behavior, i.e.~the conditional expectation, by judiciously selecting which simulations to run. This framework naturally emphasizes computational complexity and offers an abstract modular template that accommodates a variety of approximation techniques. Indeed, our template involves three major pieces: (i) experimental design to determine which simulations to run; (ii) approximation technique for the conditional expectation; (iii) optimization step to recover optimal feedback control. While these sub-problems have been treated variously elsewhere, to our knowledge we are the first ones to fully modularize and distill them in this context. In particular, emphasizing the design aspect of RMC methods was only recently taken up in \cite{gramarcy15,ludkovski15}. We borrow the design framework introduced by the first author in \cite{ludkovski15} for American options, and to our knowledge, this is the first article to explore experimental designs in context of storage problems.

We show that existing proposals for Regression Monte Carlo for storage problems all fit neatly into this template, and moreover our setup suggests a variety of further improvements. Specifically, we address the following 3 enhancements: (A) new simulation designs, including space-filling and various adaptive versions; (B) non-parametric regression architectures for learning the value function, such as Gaussian Process regression; (C) mixed approaches that vary RMC ingredients across time-steps, and even alternate between joint-$(P,I$) and discretized-inventory schemes. We emphasize the latter mix-and-match capability of our template, which allows to straightforwardly utilize different methods as $t$ changes, either deterministically or adaptively, during the backward recursion. We showcase one such possibility, addressing a non-smooth terminal condition. Mixing-and-matching subroutines is especially attractive for implementation in a broad-scope software library.

The developed Dynamic Emulation Algorithm is applicable in a wide range of stochastic storage settings, being scalable in the dimension of the state variable and potential state dynamics. We illustrate DEA with 4 different extended case-studies. The first three case-studies consider valuation of natural gas storage facilities, starting from a standard benchmark first introduced in Forsyth and Chen \cite{forsyth07}. This benchmark is then extended to add switching costs (that make control regime part of the state) and to consider simultaneously optimizing two storage facilities (leading to a 3D state space). Last but not least, we consider a rather different example from microgrid management, solving for the optimal way to dispatch backup generator in order to balance an intermittent renewable power source with a limited battery.

Our developments parallel the recent literature on \emph{optimal stopping}, especially in the context of Bermudan option pricing, where a wealth of strategies have been proposed and investigated \cite{ludkovski15}. It has been a well known folklore result that storage problems, especially with discrete controls, are ``essentially'' optimal stopping as far as computational methods are concerned. Nevertheless, the respective knowledge transfer is non-trivial and there remain substantive gaps, which we address herein, between the respective numerical algorithms. Thus, the present article ``lifts'' optimal stopping techniques to the setting of stochastic storage (i.e.~optimal switching), and can be seen as a step towards similar treatment of further stochastic control problems, such as optimal impulse, or continuous control.

The rest of the paper is organized as follows. Section \ref{sec:problem} describes the classical storage problem and the key ingredients to its solution. Section \ref{sec:numericalMethods} describes the algorithm developed to modularize the solution steps for regression Monte Carlo into a sequence of statistical learning tasks. Section \ref{sec:approximation spaces} discus the mathematics for Gaussian process regression and other popular regression methods used in the literature in the context of storage problems. In Section \ref{sec:exp design}, we introduce different design alternates such as space-filling, adaptive and dynamic to exploit the spatial information and efficiently implement non-parametric regression methods (particularly, Gaussian process regression) utilizing batched design. Sections \ref{sec:example1} and \ref{sec:microgrid} are devoted to numerical illustrations, taking up the gas storage and microgrid management, respectively. Finally, Section \ref{sec:conclusion} concludes.

\section{Problem description}
\label{sec:problem}

A storage problem is exemplified by the presence of \emph{stochastic risk factors} together with an inventory state variable. The risk factors have autonomous dynamics, while the inventory is (fully) controlled by the operator via the storage policy; thus the latter dynamics are endogenized. A second feature of storage problems we consider is their \emph{switching} property: the controller actions consist of directly toggling the storage \emph{regime}. In turn the storage regime drives the dynamics of the inventory. Depending on the setup, the regime is either a control variable, or a part of the system state.

To make our presentation concrete, we focus on the classical storage problem with a stochastic price. Namely, there are two main state variables: $P_t$ and $I_t$. $P_t \in \mathbb{R}_+$ represents the price of the stored commodity; $I_t \in [0, I_{\max}]$ is the inventory level. We  present the storage dynamics in discrete-time on a time interval $[0,T]$ discretized into a finite grid $0=t_0 < \ldots t_k \ldots < t_K = T$, such that $t_k = k\Delta t = \frac{T}{K}$.

For the price, we assume exogenous Markovian dynamics of the form
\begin{equation}
P_{\tii} = P_{\ti} + b(\ti,P_{\ti} )\Delta t + \sigma(\ti,P_{\ti} ) \, \Delta W_{\ti},
\end{equation}
where $(\Delta W_t)$ are exogenous i.i.d.~stochastic shocks. For the rest of the article we take $\Delta W_{\ti} \sim \mathcal{N}(0, \sqrt{\Delta t})$ representing Brownian motion dynamics; any other (time-dependent) shocks could be straightforwardly utilized too.
We denote by $\cF_{\ti}$ the $\sigma$-algebra generated by price process up until time $\ti$ and by $\mathbb{F} = (\cF_{\ti})$ the corresponding filtration. The inventory level $I_t$ follows
\begin{equation}
I_{\tii} = I_{\ti} + a( c_{\ti}) \,\Delta t,
\label{inventTraj}
\end{equation}
where $c_{\ti}$ is the inventory control, representing the rate of storage injection $c > 0$, withdrawal $c < 0$ or holding $c=0$. The control is linked to the storage regime $m_{\ti}$. We assume that there are three regimes $m_{\ti} \in \mathcal{J}:=\{+1,-1,0 \}$ representing injection, withdrawal and do-nothing respectively. The  regime  and control are determined by the joint state:
\begin{align}
m_{\tii} & = \cM(\ti, P_{\ti}, I_{\ti}, m_{\ti}), \label{eqn:regimeGeneral} \\
c_{\ti}(m_{\tii}) & = \cC(\ti, P_{\ti}, I_{\ti}, m_{\ti}; m_{\tii})  \label{eqn:controlGeneral}
\end{align}

Note that the above form  implies that at each time-step $\ti$ and state $(P_\ti,I_\ti,m_\ti)$ the controller picks her next regime $m_\tii$ which in turn determines her control $c_\ti (m_\tii) $. It also directly restricts controls to be of Markovian feedback form, making the policies $(c_{\ti},m_{\tii})$ $(\cF_{\ti})$-adapted. As a result, $(P_{\ti}, I_{\ti}, m_{\ti})$ is a Markov process, adapted to the price filtration $(\cF_{\ti})$.

Let $\pi(P,c)$ be the instantaneous profit rate earned by using control $c$ when price is $P$, and $K(i, j) \ge 0$ be the switching cost for switching from regime $i$ to $j$. Then
$$\pi^{\Delta}(P_\ti,m_\ti, m_\tii) := \pi(P_{t_k},c_\ti(m_\tii))\Delta t - K(m_\ti, m_\tii),$$
is  the net profit earned during one time-step $[\ti, \tii )$. To denote the cumulative profit of the controller on $[\ti,T]$ along the path specified by $\textbf{P}_{\ti} = P_{t_k:t_K}$ and a selected sequence of regimes $\textbf{m}_\ti := (m_{t_k:t_K})$  (and consequently the control $\textbf{c}_\ti := (c_{t_k:t_K})$) we use 
\begin{equation}
v(\ti,\textbf{P}_{\ti},I_\ti,\textbf{m}_{\ti}) := \sum_{s=k}^{K-1} e^{-r(\ts-\ti)} \pi^{\Delta} (P_\ts,m_\ts,m_\tss) + e^{-r(T-\ti)}W(P_T,I_T),
\label{eqn:pathwise}
\end{equation}
where $r \ge 0$ is the discount rate and $W(P,I)$ is the terminal condition (typically concerning the final inventory $I$) at the contract expiration. Note that $I_T$ is determined recursively based on $\textbf{m}_\ti$ using \eqref{inventTraj}. The goal of the controller is to maximize discounted expected profits on the horizon $[\ti,T]$
\begin{align}
\label{objFn}
V(\ti,P,I, m) & = \sup_{\textbf{m}_{\ti} } \mathbb{E} \left[ v(\ti,\textbf{P}_{\ti},I_{\ti}, \textbf{m}_{\ti}) \Big| \ P_\ti = P,I_\ti = I, m_\ti = m \right], \\
& \text{subject to }
  I_{t_s} \in [I_{min},I_{max}] \quad \forall s. \label{cond_dt}
\end{align}

Problem \eqref{objFn} belong to the class of  stochastic optimal control, and satisfies the dynamic programming principle. Namely,
\begin{equation}
V(t_k,P_{t_k},I_{t_k},m_{t_k}) = \max_{m \in \mathcal{J} } \mathbb{E} \left[\pi^{\Delta}(P_\ti,m_\ti,m) + e^{-r \Delta t}V(\tii, P_{\tii},I_{\tii}, m)  \Big| P_{t_k} \right],
\label{DPP_discret1}
\end{equation}
where the expectation is over the random variable $P_\tii$ since the inventory $I_\tii$ is fully determined by $I_\ti$ and the chosen regime $m$ on $[\ti, \tii)$.
Note that due to the inventory constraints, some of the regimes might not be admissible for different initial conditions, so that formally the maximum in \eqref{DPP_discret1} is over $\mathcal{J} = \mathcal{J}(\ti,P_\ti,I_\ti, m_\ti) \subseteq \{+1, -1, 0\}$. For instance, if the inventory is zero, $I_\ti = 0$ further withdrawal is ruled out and $\mathcal{J}(\ti,P_\ti,0,m_\ti) = \{0,1\}$. Such constraints could even be time- or price-dependent, for instance in hydropower management.

\begin{remark}
  In our main setup the choice of the control is pre-determined given the stochastic state and the regime. More generally,  conditional on the regime there might be a set of admissible controls $\cA(\ti,m_{\tii})$ adding an additional optimization sub-step. For example, we might have $\cA(\ti,+1) = (0,c_{\max}(I_\ti) ]$ and $\cA(\ti,-1) = [c_{\min}(I_\ti),0 )$, where $c_{\min}$ is the maximum withdrawal rate and $c_{\max}$ is the maximum injection rate. When $|\cA|>1$, the optimization problem \eqref{DPP_discret1} requires first to find the optimal regime $m_\tii$, and secondly to find the optimal control $c_\ti(m_\tii)$ admissible to this regime.
    The original situation then corresponds to $\cA$ being a singleton, and can be interpreted as a trivial optimization over $\cA$ e.g.~due to a bang-bang structure. Abstractly we may always write $c_\ti(m_\tii)=\cC(\ti, P_{\ti}, I_{\ti}, m_{\ti}; m_{\tii})$ subsuming the inner optimizer if necessary.
\end{remark}

\subsection{Solution Structure}
\label{solution}
Due to the Markovian structure, at each time-step $\ti$ and state $(P_\ti,I_\ti,m_\ti)$ the controller picks her next regime $m_\tii$ (consequently control $c_\ti (m_\tii)$) 
\begin{equation}
m^*(t,P,I,m) = \argmax_{j \in \mathcal{J}}  \{  \pi^{\Delta}(P,m,j) + q(t,P,I+a(c_t(j))\Delta t,j ) \}
\label{controlMap}
\end{equation}
where
\begin{equation}
q(\ti,P,I,m):=\mathbb{E} \left[e^{-r \Delta t} V(\tii, P_\tii,I, m ) \big| \, P_\ti =P \right].
\label{contVal}
\end{equation}
Conceptually, we have a map from the value function $V$ to the \emph{chosen} $m^*$ and $c^*$, encoded as $m^* : (t,P,I,m) \mapsto \mathcal{J}$. The explicit dependence of the continuation value $q$, value function $V$ and control map $m^*$  on the regime $m$ is a consequence of switching costs $K(m_\ti, m_\tii)$ whose absence will result in a lower dimensional problem to solve, see Remark \ref{rem:dim-reduce}.

Conversely, equation \eqref{objFn} provides the representation of the value function as a conditional expectation of future profits based on an optimal policy $\mathbf{m}^*$. Therefore, any estimate $\hat{m}: (t,P,I,m) \mapsto \mathcal{J}$ of the control map naturally induces a corresponding estimate $\hat{V}$ of the value function. Specifically, $\hat{m}$ yields the dynamics
$$
\hat{I}_\tii = \hat{I}_\ti + a\big({c}_\ti(\hat{m}_\tii(t,P,I,\hat{m}_\ti)) \big) \Delta t,
$$
which in turn can be used for out-of-sample forward simulations, $$\hat{V}(0,P_0,I_0,m_0) = \E \left[
\sum_{s=1}^{K-1} e^{-r \ts} \pi^{\Delta} (P_\ts,\hat{m}_\ts,\hat{m}_\tss) + e^{-r T}W(P_T,\hat{I}_T) \right].$$
While $\hat{I}_\ts$ does not appear explicitly above, it is crucially driving $\hat{m}_\tss(t,P_\ts,\hat{I}_\ts,\hat{m}_\ts)$.

\begin{figure}[ht]
  \centering
    \includegraphics[scale=0.4]{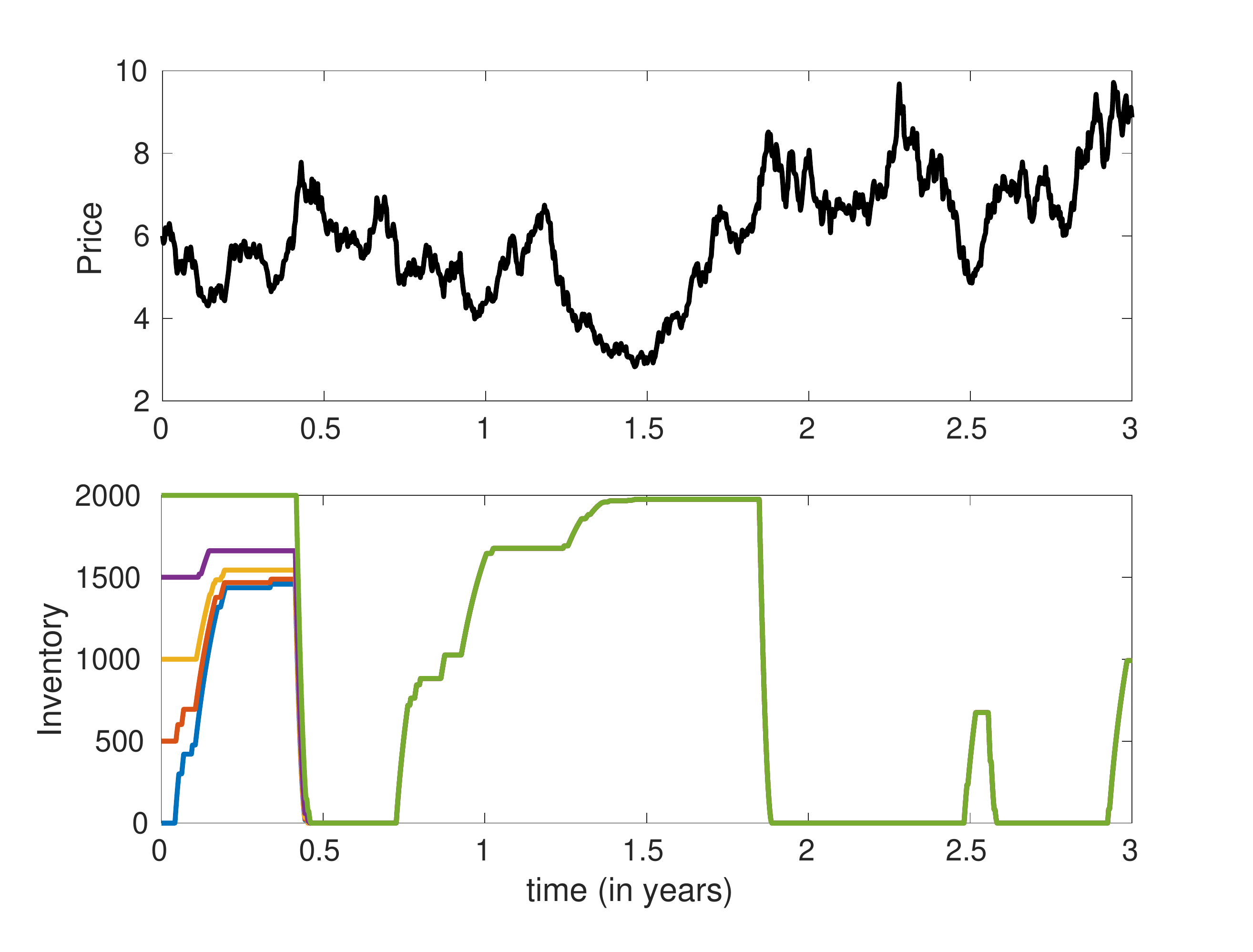}
\caption{\textit{Top panel:} a given trajectory of commodity price $(P_t)$ following logarithmic mean reverting process in~\eqref{eqn:priceProcess}.  \emph{Lower panel:} Corresponding trajectories of controlled inventory $\hat{I}_t$ starting at $\hat{I}_0 \in \{0,500,1000,2000\}$. This figure is associated with the gas storage example of Section \ref{sec:storage-1} using the PR-1D solution scheme. \label{fig:trajectory}}
\end{figure}

\begin{figure}[!ht]
  \centering
     \begin{subfigure}[b]{0.3\textwidth}
    \includegraphics[width=\textwidth]{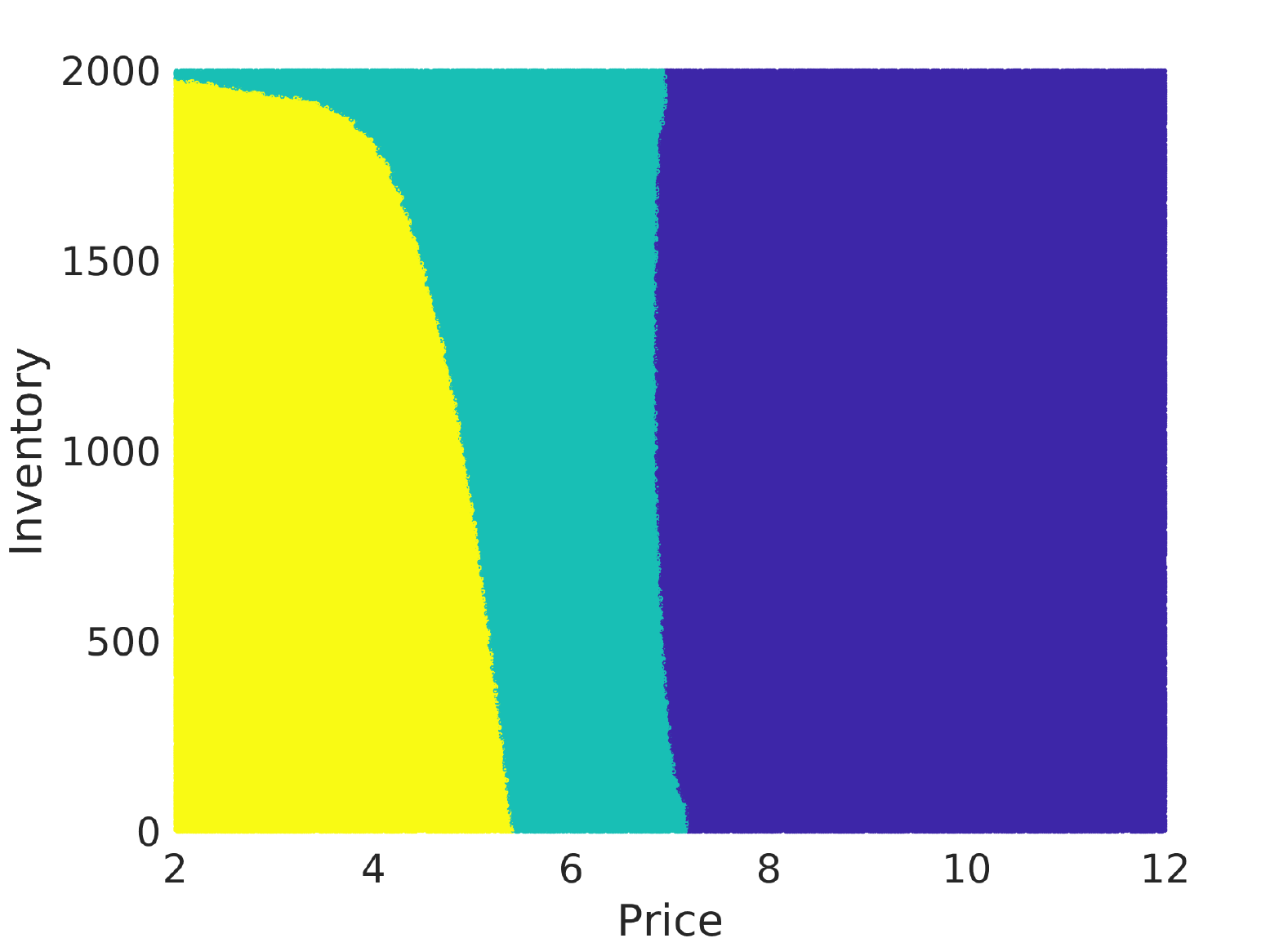}
    \caption{Control map}
    \label{fig:controlMapEvolution_b}
  \end{subfigure}
 \quad
 \begin{subfigure}[b]{0.3\textwidth}
    \includegraphics[width=\textwidth]{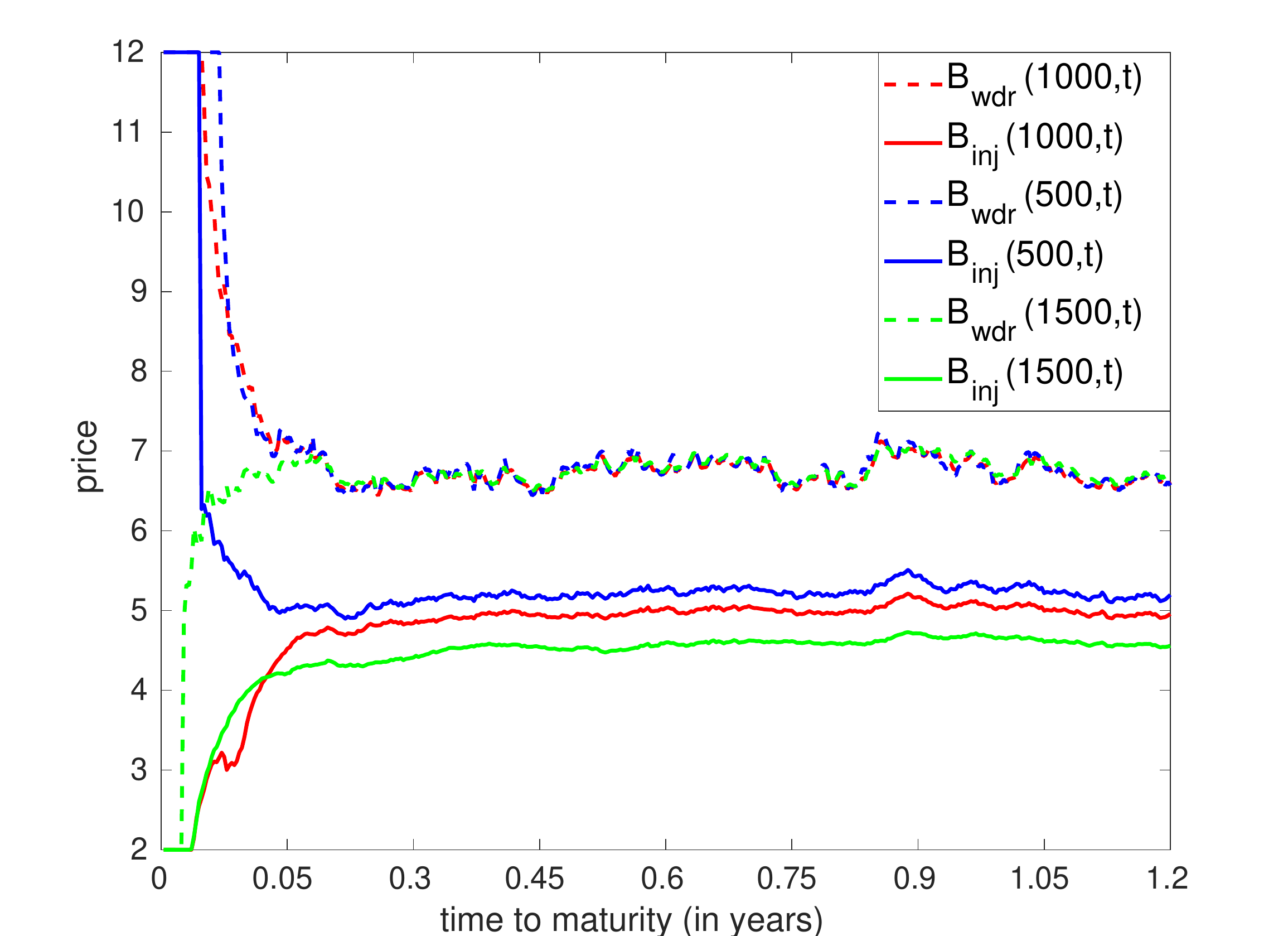}
    \caption{\small{Prod./Inject. boundaries}}
    \label{fig:controlMapEvolution_a}
  \end{subfigure}
 \quad
    \begin{subfigure}[b]{0.3\textwidth}
    \includegraphics[width=\textwidth]{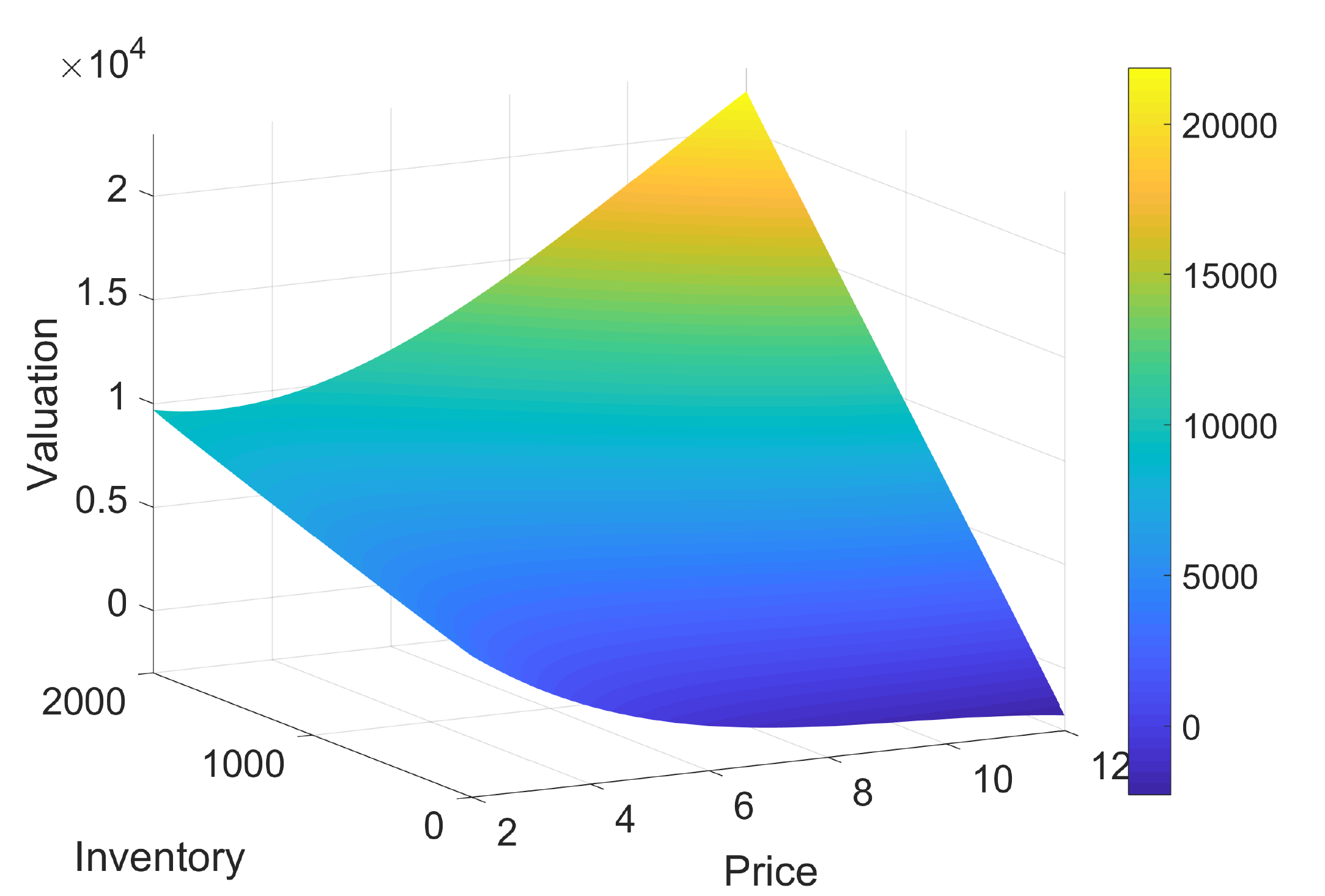}
    \caption{Value function}
    \label{fig:3dvaluefunction}
  \end{subfigure}
\caption{\textit{Left panel:} snapshot of the control map  $\hat{m}(t,P,I)$ at $t=2.7$ years. \emph{Middle:}  injection $B_{inj}(I,\cdot)$ and withdrawal $B_{wdr}(I,\cdot)$ boundaries as a function of time for $I \in \{500, 1000, 1500\}$. \emph{Right:} value function $\hat{V}(0,P,I)$ at $t=0$. Results were obtained with conventional design and PR-1D regression.}
\label{fig:controlMapEvolution}
\end{figure}

Figure \ref{fig:trajectory} illustrates this dual link by showing a trajectory of $(P_t)$ and several corresponding trajectories  of $(\hat{I}_t)$  indexed by their initial inventories ${I}_0$ (viewed as an external parameter) for a gas storage facility (see section \ref{sec:storage-1} for more details). One interesting observation is that the dependence of time-$t$ inventory $\hat{I}_t$ on ${I}_0=i$ is rather weak, i.e.~the inventory levels coalesce:  $\hat{I}^{i}_t = \hat{I}^{i'}_t$ after an initial ``transient'' time period. Note that because the controls are specified in feedback form, once $\hat{I}^{i}_t = \hat{I}^{i'}_t$ we have $\hat{m}^i_\ts = \hat{m}^{i'}_\ts$ and the inventory paths will stay together forever. The Figure also illustrates the underlying maxim of ``buy low, sell high'': when $P_t$ is low, controlled inventory $\hat{I}_t$ is high (and increasing), and when $P_t$ is high, $\hat{I}_t$ is low (and shrinking). As a result, we see a clustering of $\hat{I}_t$ around the minimum and maximum storage levels $I_{min}, I_{max}$, indicating the strong constraint imposed by  the bounded storage capacity.
To visualize the estimated optimal policy $\hat{m}$, Figure~\ref{fig:controlMapEvolution_b} plots the control map $ (P,I) \mapsto \hat{m}(t, P, I)$ at a fixed time step $t$, namely with $T-t=0.3$ years for the gas storage example of section \ref{sec:storage-1}. The state space is divided into three regions: when $P$ is high it is optimal to withdraw: $\hat{m} = -1$; if $P$ is low, it is optimal to inject $\hat{m} = +1$; in the middle, or if inventory is very large, it is optimal to do nothing $\hat{m} = 0$. Typically, this map is interpreted by fixing current inventory $I$ and looking at $\hat{m}$ as a function of $P$. We can then summarize the resulting policy in terms of the \emph{injection/withdrawal boundaries} $B_{inj}(I,t)$ and $B_{wdr}(I,t)$:
\begin{align}
    B_{inj}(I,t) & := \sup \{P_t: \hat{m}(t,P_t,I)=+1  \}, \notag \\
    B_{wdr}(I,t) & := \inf \{P_t: \hat{m}(t,P_t,I)=-1  \}.
    \label{defn: injectionWithdrawal}
\end{align}
Since injection becomes profitable with low prices, $B_{inj}(I,t)$ represents the maximum price for which injection ($\hat{m}(t,P_t,I)=+1$) is the optimal policy. Similarly, $B_{wdr}(I,t)$ represents the minimum price for which  withdrawal ($\hat{m}(t,P_t,I)=-1$) is the optimal policy. The interval $[B_{inj}(I,t), B_{wdr}(I,t)]$ is the no-action region. These boundaries are plotted as a function of $T-t$ in Figure~\ref{fig:controlMapEvolution_a} at three different inventory levels. One prominent feature is the boundary layer as $T-t \to 0$ whereby the policy is primarily driven by the terminal penalty $W(P,I)$ than immediate profit considerations. In this example, due to the ``hockey-stick" penalty $W(P_T,I_T)$ which forces the controller to target the inventory level $I=1000$, as $T-t \to 0$, injection becomes the optimal policy for $I<1000$ independent of the price, and withdrawal becomes optimal for $I>1000$ (the no-action region effectively disappears). Conversely, for large $T-t$, the boundaries $t\mapsto B_{inj}(I,t)$ and $t \mapsto B_{wdr}(I,t)$ are essentially time-stationary.

Finally, Figure~\ref{fig:3dvaluefunction} shows the value function $\hat{V}(0,P,I)$ as a 2-D surface in the price and inventory coordinates. As noted by previous studies, for a fixed price $P$, we observe a linear relationship between value and inventory. However, as a function of $P$, $V(0,\cdot,I)$ is non-linearly decreasing at low inventory levels, and increasing for large inventory.

\section{Dynamic Emulation Algorithm}
\label{sec:numericalMethods}

The numerical algorithms we consider provide approximations, denoted as $\hat{q}(\cdot)$,  to the continuation value in \eqref{contVal}.
We then obtain a recursive solution for $V(\cdot)$ through backward induction on $\ti$.  Namely, starting with the known terminal condition
$$
V(T,P_T,I_T,m) = q(t_K,P_{t_K},I_{t_K},m) =W(P_T,I_T)  \quad \forall m,
$$
we apply backward induction to estimate $\hat{q}(\ti, \cdot, \cdot, \cdot)$ as $k=K -1$ to $k = 0$.
 The recursive construction ensures that at step $t_k$ we know the continuation functions $\hat{q}(\tii, \cdot, \cdot, \cdot)$ and hence can find $\hat{m}(\tii,P,I,m)$ as in \eqref{controlMap}. We then  take
 \begin{align}
   \hat{V}(\ti,P,I,m) = \pi^\Delta(P,m,\hat{m}_\tii) + \hat{q}(\tii,P,I+a(c(\hat{m}_\ti))\Delta t, \hat{m}_\tii)
 \end{align}
 and complete the inductive step by learning $\hat{q}(\ti,\cdot)$ as the conditional expectation of $e^{-r \Delta t}\hat{V}(\tii, \cdot)$.

\begin{remark}
\label{re:look-ahead}
Above we view the continuation value as the conditional expectation of one-step-ahead value function, matching the original \cite{tvr} scheme. More generally, we can unroll the Dynamic Programming equation \eqref{DPP_discret1} using the optimal regime choices $m^*_{\tii}$ and corresponding controls $c^*_{t_k}$ for any $w \ge 1$ as
\begin{align}\notag
V(\ti,P_\ti,I_\ti, m_\ti) & =\mathbb{E}[  \pi^{\Delta}(P_\ti,m_\ti,m_\tii^*) + q(\ti,P_\ti,I_\tii,m_\tii^*) \Big| P_\ti ]\\ \notag
 =\mathbb{E}\Big[  \pi^{\Delta} (P_\ti,m_\ti,m_\tii^*) &  + e^{-r \Delta t}\pi^{\Delta}(P_\tii,m_\tii^*,m_{t_{k+2}}^*) + e^{-r \Delta t}q(\tii,P_\tii,I_{t_{k+2}}^*,m^*_{t_{k+2}}) \Big| \, P_\ti \Big]\\
& \qquad \ldots  \notag \\
& = \mathbb{E} \Bigl[v_{{k:k+w}}(\pi,q)(\textbf{P}_\ti,\textbf{I}_\ti, \textbf{m}_\ti) \Big|\, P_{t_k},I_{t_k},m^*_{t_k}\Bigr]
 \label{valuFn_discrete}
\end{align}
in terms of the pathwise gains
\begin{multline}
v_{{k:k+w}}(\pi,q)(\textbf{P}_\ti,\textbf{I}_\ti,\textbf{m}_\ti) := \sum_{s=k}^{k+w-1} e^{-r (s-k)\Delta t} \pi^{\Delta}(P_{t_s},m_{t_s},m_\tss^*) \\ +e^{-r w \Delta t}q(t_{k+w},P_{t_{k+w}},I_{t_{k+w+1}},m_{t_{k+w+1}}).
\end{multline}
Similarly,  the continuation value $q(\ti,P_{\ti},I_{\tii},m_{\tii})$ can be written as
\begin{align}
\label{eqn: contFnANdValFn}
q(\ti,P_\ti,I_\tii,m_\tii) &= \mathbb{E} \Bigl[v_{{k+1:k+w}}(\pi,q)(\textbf{P}_\tii,\textbf{I}_\tii,\textbf{m}_\tii) | P_\ti,I_\ti,m_\ti\Bigr].
\end{align}
The partial-path construction in \eqref{eqn: contFnANdValFn} which can be traced back to \cite{egloff05,egloff07}. It encompasses the Tsitsiklis-Van Roy (TvR) algorithm \cite{tvr} where $w=1$, and the Longstaff-Schwartz (CLS) algorithm \cite{ls2001} where $w = K - k -1$.  Through the rest of this paper, we will fix $w=1$ and use short-hand notation for $v_{{k:k+1}}(\pi,q)(\textbf{P}_\ti,\textbf{I}_\ti,\textbf{m}_\ti) $ as $v_{k}(\textbf{P}_\ti,\textbf{I}_\ti,\textbf{m}_\ti) $ or just $v_{k} $.
\end{remark}

 The simulation-based framework relies on generating  pathwise profits $v_{k+1}(\textbf{P}_\tii,\textbf{I}_\tii,\textbf{m}_\tii)$ and recovering the continuation function $\hat{q}(\ti,\cdot) $ as its expected value as in~\eqref{eqn: contFnANdValFn} via a Monte Carlo approximation. Namely, one aims to   $L^2$-project $v_{{k+1}}$ onto an approximation space $\mathcal{H}_k$ using a \emph{regression} procedure:
\begin{equation}
  \check{q}(\ti, \cdot) := \argmin_{h_\ti \in \mathcal{H}_k} \| h_\ti - v_{{k+1}} \|_2.
\end{equation}
As a canonical setup, $\mathcal{H}_k = \spn( \phi_1, \ldots, \phi_R)$ is the linear space generated by basis functions $\phi_i$ and so the approximation $\check{q}(\ti,\cdot) = \sum_{i=1}^R \beta_i \phi_i(\cdot)$ is described through its coefficient vector $\vec{\beta}$.

To estimate $\vec{\beta}$ we solve a discrete optimization problem based on a simulation design $\mathcal{D}_k :=(P^n_\ti, I^n_\tii)_{n=1}^N$ and the corresponding realized pathwise values $v^n_{{k+1}}, n=1,\ldots, N$ from trajectories started at $(P^n_{\ti}, I^n_{\tii}, m_{\tii})$:
\begin{equation}
\hat{q}(\ti, \cdot, \cdot, m_\tii) = \argmin_{h_\ti \in \mathcal{H}_k} \sum_{n=1}^{N} |h_\ti(P_\ti^n,I_\tii^n) - v^n_{t_{k+1}}|^2.
\label{eqn:RNMC}
\end{equation}
Thus, $\hat{q}$ is an empirical approximation of the projection $\check{q}$, with the corresponding finite-sample error. In particular,
subject to mild conditions on $\CD_k$, we have $\hat{q} \to \check{q}$. Traditionally \eqref{eqn:RNMC} is implemented by generating $N$ global paths $(P^n_{t})$ for the exogenous (price) process starting at $t=0$ until maturity $T$, which are then stored permanently in memory for the entire  backward induction, introducing significant overhead. Algorithm  \ref{algo_Generalized} replaces this with a design $\CD_k$ and the associated one-step trajectories $(P^n_{t:t+1}, I^n_{t+1})$.

\begin{remark}\label{rem:dim-reduce}
In the case where there are no switching costs $K(i,j) \equiv 0 \ \forall i,j$, the dimensionality  of the regression problem can be reduced from 3 to 2. Indeed, since the inventory process $I_t$ is completely controlled, the continuation value only depends on the inventory and regime $(I_\ti, m_\ti)$ through the next-step inventory $I_\tii =  I_\ti +a(c(m_\ti) )\Delta t$. Analogously, the value function is then independent of the current regime, $V(\ti, P_\ti, I_\ti)$.  With a slight abuse of notation we then write $q(\ti, P_{\ti}, I_{\tii})$, working with the projection subspace generated by $(P_\ti,I_\tii)$.

When switching costs are present, the same reduction is possible during the regression, but the present regime $m_\ti$ remains a part of the state since it affects the continuation value $q$, so no overall dimension savings are achieved. Practically, this is handled by solving a distinct regression for each $m \in \mathcal{J}$ as in \eqref{eqn:RNMC}.
\end{remark}

To apply \eqref{eqn:RNMC}, we furthermore need to evaluate continuation values at arbitrary future states $(P^n_{t_s}, I^n_{t_s}, m^n_{t_s})$, $s > k$ which corresponds to predicting the previous $\hat{q}(\ts, \cdot, \cdot, \cdot)$. These two operations---\texttt{fit} and \texttt{predict}---are the main workhorses of the statistical approximation procedure.

 The above procedure generates the estimated continuation values $\hat{q}(\ti, \cdot, \cdot, \cdot)$, the corresponding estimated optimal regime $\hat{m}(\tii, \cdot, \cdot, \cdot)$  matching \eqref{controlMap}, and consequently the control ${c}(\hat{m})$.
  Finally, the value function $V(0,P_0,I_0,m_0)$ at $t_0=0, P_0, I_0, m_0$ is approximated \emph{out-of-sample} by generating $N'$ sample paths $(P^{n'}_{0:T}, \hat{I}^{n'}_{0:T}, m^{n'}_{0:T}),$ where the pathwise inventory $\hat{I}$ is based on the just-estimated optimal control map $\hat{m}$, cf.~Figure \ref{fig:trajectory}. Thus, $\hat{V}(0,P_0,I_0,m_0) = \frac{1}{N'} \sum_{n'=1}^{N'} v_{0:T}(\pi,\hat{q})(0,\mathbf{P}_0^{n'},\mathbf{I}_0^{n'},\mathbf{\hat{m}}_0^{n'})$, where $v_{0:T}(\pi,\hat{q})(0,\mathbf{P}_0^{n'},\mathbf{I}_0^{n'},\mathbf{\hat{m}}_0^{n'})$ is the estimated total cumulative discounted profit using $\hat{m}$ for each sample path. Since the policy $\mathbf{\hat{m}}$ is necessarily sub-optimal, $\hat{V}(0,\cdot)$ is a lower bound for the true $V$, modulo the Monte Carlo error from the $N'$ trajectories used in the last averaging.
  \begin{remark}
    Note that the final estimate is formally a function of the out-of-sample simulations $P^{n'}_{0:T}$, as well as the in-sample simulations $(P^{n}_{t:t+1})$. To facilitate comparison of different methods, where possible we fix the ``test'' scenario database $(P^{n'}_{0:T})$, whereby we can directly evaluate the different controls/cumulative revenues obtained along a given sample path of the price process.
  \end{remark}

The Dynamic Emulation  Algorithm  \ref{algo_Generalized} (DEA) presents the overall template for solving the storage problem. The chosen nomenclature is because the algorithm allows the user to change the projection space $\cH_k$ and simulation designs  $\CD_k$ with time, hence it is ``dynamic''; and ``emulation'' to reflect the statistical learning perspective that targets the approximation or emulation of the continuation value $q(t,\cdot)$. Its overall complexity is $\mathcal{O}(KN)$.  Relative to existing literature, DEA adds the following tools:

\begin{itemize}
\itemsep-0.3em
\item Possibility to include different regression spaces $\cH_k$ across timesteps $t_k$;
\item General simulation designs $\CD_k$, again potentially varying across time-steps;
\item Eliminating the requirement to store global price paths in memory, rather new paths are simulated ``online", i.e., at every time step. This also allows to vary the number of simulations $N_k$ at time step $t_k$. 
\end{itemize}

\RestyleAlgo{boxruled}
\LinesNumbered
\begin{algorithm}[!ht]
\SetAlgoLined
\KwData{$K$ (time steps), $(N_k)$ (simulation budgets per step) }
Generate design $\CD_{K-1,m} := (\bP_{K-1}^{\mathcal{D}_{K-1,m}},\bI_K^{\mathcal{D}_{K-1,m}})$ of size $N_{K-1}$ for each $ m \in \mathcal{J} $. \\
Generate one-step paths $P_{K-1}^{n,\mathcal{D}_{K-1,m}} \mapsto P_{K}^{n,\mathcal{D}_{K-1,m}}$ for $n=1,\ldots,N_{K-1}$ and $ m \in \mathcal{J} $\\
Terminal condition:   $v_{K,m}^n \leftarrow W(P_{K}^{n,\CD_{K-1,m}},I_{K}^{n,\CD_{K-1,m}})$ for $n=1,\ldots, N_{K-1}$ and $ m \in \mathcal{J} $\\
\For{$ k=K - 1, \ldots, 1$ }{
\For{$ m \in \mathcal{J} $ }  {
$\hat{q}(k,\cdot,\cdot,m) \leftarrow \arg \min_{h_k \in \mathcal{H}_k} \sum_{n=1}^{N_k} |h_{k}(P_{k}^{n,\mathcal{D}_{k,m}},I_{k+1}^{n,\mathcal{D}_{k,m}}) - v_{k+1,m}^n |^2$  \\
Generate design $\CD_{k-1,m} := (\bP_{k-1}^{\mathcal{D}_{k-1,m}},\bI_{k}^{\mathcal{D}_{k-1,m}})$ of size $N_{k-1}$ for each $ m \in \mathcal{J} $ \\
Generate one-step paths $P_{k-1}^{n,\mathcal{D}_{k-1,m}} \mapsto P_{k}^{n,\mathcal{D}_{k-1,m}}$ for $n=1,\ldots,N_{k-1}$\\
}
\For{$n=1,\ldots,N_{k-1}$ \text{and} $ m \in \mathcal{J} $  }  {
$m' \leftarrow \argmax_{j \in \mathcal{J}} \{ \pi^\Delta(P_{k}^{n,\mathcal{D}_{k-1,m}},m, j) + \hat{q}(k,P_{k}^{n,\mathcal{D}_{k-1,m}},I_{k}^{n,\mathcal{D}_{k-1,m}}+a(c_{k}(j))\Delta t, j)\}$ \\
	$v^n_{k,m} \leftarrow  \pi^\Delta(P_{k}^{n,\mathcal{D}_{k-1,m}},m,m') + e^{-r \Delta t} \hat{q}(k,P_{k}^{n,\mathcal{D}_{k-1,m}},I_{k}^{n,\mathcal{D}_{k-1,m}}+a(c_{k}(m'))\Delta t, m')  $
}

}
return $\{ \hat{q}(k,\cdot,\cdot,m) \}_{k=1,m \in \mathcal{J}}^{K-1}$\\
\caption{Dynamic Emulation Algorithm (DEA) - $\mathcal{O}(KN)$}
\label{algo_Generalized}
\end{algorithm}

An alternative to \eqref{eqn:RNMC} is to use as state variables $(P_\tii, I_\tii)$ during the projection, and then take the conditional expectation analytically:
\begin{equation}
\label{eqn:RLMC}
\begin{aligned}
\hat{q}(\ti, P, I, m) &= \mathbb{E}\left[ \displaystyle\argmin_{h_\tii \in \mathcal{H}_{k+1}} \sum_{n=1}^{N} |h_\tii(P_\tii^n,I_\tii^n) - v^n_{{k+1}}|^2 \Big| \ P_\ti=P, I_\tii=I \right].
\end{aligned}
\end{equation}
This is known as Regress Later Monte Carlo (RLMC) and lowers the variance in the estimated $\hat{q}$ \cite{nadarajah17}. However, the requirement of closed form expressions for the conditional expectation generally rules out use of RLMC with non-parametric approximation spaces.

The following two sections provide menus for the two principal steps in DEA: selecting the approximation space $\mathcal{H}_k$ and the design $\CD_k$. Note that these can be mixed and matched across $\ti$.  In aggregate, the two steps correspond to a machine learning task: given a stochastic simulator (that returns one-step-ahead profits given initial condition $(t,P,I,m)$), we wish to learn the input-output relationship, i.e.~to predict the expected response for any (in-sample, or a new out-of-sample) input.
Therefore, in that language the DEA is a (recursive) sequence of learning tasks, with the performance measure given by the quality of the final answer $\hat{V}(0,\cdot, \cdot, \cdot)$. Finally, we conclude this section with a comment on another attribute of the algorithm, namely look-ahead parameter $w$ which we don't explore in this work.

\section{Approximation Spaces}
\label{sec:approximation spaces}
In this section we fix a given time step $\ti$ and consider the problem of approximating the continuation value $q(\ti_, P, I, m)$ viewed as a function $h_\ti(P,I)$. Below we generically use $\bx = (P_\ti^n,I_\tii^n)_{n=1}^N$ and $\by = (v^{n}_{k+1})_{n=1}^{N}$ to represent the dataset used during the regression. We note that while the full state space is in general $(P, I, m)$, since $m$ is a factor variable (rather than $\mathbb{R}$-valued), it is handled discretely, i.e.~a separate approximation $\hat{h}(\cdot,m)$ is constructed for each level of $m \in \mathcal{J}$ (so three separate approximations in the typical storage setting). The latter issue is moot when there are no switching costs and the dimensionality of the continuation value is only 2.

The statistical assumption is that the input-output relationship is described by
\begin{equation}
    y^n = h(x^n) + \sigma^2\xi, \ \text{where } \xi \sim \mathcal{N}(0,1),
\end{equation}
where $h \in \mathcal{H}_k$ is the unknown function to be learned, and $\sigma^2 \xi$ is the noise. In our case, the noise is due to the stochastic shocks in $(P_\tii)$ relative to $P_\ti$, which induce variation in the realized pathwise profit relative to the average continuation value.

The classical regression framework is a linear model where $\mathcal{H}_k$ is the vector space spanned by some basis functions $(\phi_i)$. Then the prediction at a generic $x_*$ is controlled by the regression coefficients $\vec{\beta}$: $h(x_*) = \vec{\beta}^T \vec{\phi}(x_*)$.  The key challenge is then to specify the basis functions $\phi_i$, because the intrinsic approximation error (i.e.~the distance between the true $q(\ti,\cdot)$ and $\mathcal{H}_k$) strongly affects the quality of the solution. As a result, various regression methods have been explored in the literature, for example, global polynomial regression \cite{balata17,boogert08,ludkovski10}, radial basis functions \cite{boogert13}, support vector regressions \cite{malyscheff17}, kernel regressions \cite{langrene15}, neural network \cite{han16} and piecewise linear regression \cite{warin}. Inspired by the recent work \cite{ludkovski15}  on Bermudan options, below we additionally introduce the usage of Gaussian process (GP) regression  for solving storage problems. To our knowledge, ours is the first paper to use GPs in such context.

 \subsection{Bivariate piecewise approximation}
 \label{regMethod}
Following Longstaff-Schwartz's seminal work~\cite{ls2001} on Bermudan option pricing, researchers have employed polynomial regression (PR) also for storage problems \cite{balata17,boogert08,ludkovski10}. A degree-$r$ global polynomial approximation $h_\ti = \sum_i \beta_i \phi_i$,
has $(r+1)(r/2+1)$ basis functions and takes $\phi_i(P,I) = P^{\alpha_1(i)} I^{\alpha_2(i)}$, where the total degree of the basis function is $\alpha_1 + \alpha_2 \le r$. For example, a global quadratic approximation ($r=2$) has 6 basis functions $\{1, P, P^2, I, I^2, P\cdot I\}$, and a cubic PR has 10 basis functions. Our experiments indicate that typically PR leads to poor performance due to the resulting stringent constraints on the shape of the continuation value and consequent back-propagation of error.

A popular alternative is to use piecewise approximations based on partitioning the space of $(P,I)$, restricted to $[\min_{1 \le n \le N} P_{t_i}^n, \max_{1 \le n \le N} P_{t_i}^n] \times [I_{\min}, I_{\max}]$, into $M = M_P \times M_I$ rectangular sub-domains $\CD_{i_1,i_2}$, $i_1 = 1,2,\ldots,M_P; i_2 = 1,2,\ldots,M_I$. We then consider basis functions of the form $\lbrace \phi_g^{i_1,i_2}\rbrace$, with support restricted to $\CD_{i_1,i_2}$. For example, we can take piecewise linear approximation, i.e.~$g=1,2,3$ with
\begin{align*}
\phi_1^{i_1, i_2}(P,I) &= \mathbf{1}_{ (P,I) \in \CD_{i_1,i_2} } \\
\phi_2^{i_1, i_2}(P,I) &= P \cdot \mathbf{1}_{ (P,I) \in \CD_{i_1,i_2} } \\
\phi_3^{i_1, i_2}(P,I) &= I \cdot \mathbf{1}_{ (P,I) \in \CD_{i_1,i_2} }.
\end{align*}
Overall we then have $3 M_P M_I$ coefficients to be estimated.
Higher-degree terms could also be added, e.g.~a cross-term $P \cdot I$ or quadratic terms $P^2, I^2$. Piecewise regression offers a divide-and-conquer advantage with the overall fitting done via a loop across $\CD_{i_1,i_2}$'s; in each instance only a small subset of the data is selected to learn a few coefficients. This decreases the overall workload of the regression substep, and allows for parallel processing. Relative to PR, piecewise regressions are also more ``robust'' to fitting arbitrary shapes of the continuation value. Their main disadvantage is the inherent discontinuity in $\hat{q}$ at the sub-domain boundaries, and the need to specify $M_P, M_I$  and then construct the rectangular sub-domains $\CD_{i_1,i_2}$. We refer to \cite{warin} for several adaptive constructions, including equal-weighted and equi-gridded.

\subsubsection{Piecewise continuous approximation}\label{sec:piecewise}
Another approach is to construct a piecewise approximation that is continuous in one dimension through a linear interpolation. For example, after discretizing the endogenous inventory variable into $M_I+1$ levels $I_0,I_1,\ldots,I_{M_I}$, we fit an independent degree-$M_P$ monomial in $P$ for each level, i.e.~optimize for $\hat{h}_j(P) := \sum_i \beta_{ij} \phi_i(P)$ for $j=0,\ldots, M_I$, giving  a total of $(M_I+1)M_P$ regression coefficients.
(Conversely, one could also fit a piecewise linear model with $M_P$ sub-domains in the $P$-dimension.) The final interpolated prediction for arbitrary $I \in (I_j,I_{j+1})$ is then piece-wisely defined as
\begin{equation}
   \widehat{h}_{\ti}(P,I) := \delta(I) \ \hat{h}_{j}(P) + (1-\delta(I)) \ \hat{h}_{j+1}(P),
   \label{eqn:interpol}
\end{equation}
where $\delta(I)  = \frac{I-I_{j}}{I_{j+1} - I_{j}}$. 

\subsection{Local polynomial regression (LOESS)}
Local regressions minimize the worry regarding the choice of basis functions by constructing a non-parametric fit that
solves an optimization problem at each predictive site. Given a dataset $\{ \bx, \by\}$ (as a reminder, $x \equiv (P,I)$ is 2-dimensional throughout this section), the prediction using LOESS at $x_*$ is $h(x_*) = \sum_{i=1}^r \beta_i(x_*) \phi_i(x_*)$
where the \emph{local} coefficients $\beta_i(x_*)$ are determined from the weighted least-squares regression
\begin{align}
    \vec{\beta}(x_*)  &= \argmin_{\vec\beta \in \mathbb{R}^r}{\frac{1}{N}\sum_{n=1}^{N}\kappa(x_*,{x}^n)\left[y^n - \vec{\beta}^T \vec{\phi}(x^n)
    \right]^2}.   \label{eqn:loess}
\end{align}
The weight function $\kappa(x_*,x) \in [0,1]$ gives more weight to $y^n$'s from inputs close to $x_*$, akin to kernel regression. 
Since a separate optimization is needed for each  $x_*$, to make $N'$ predictions \eqref{eqn:loess} has complexity of $\mathcal{O}(NN')$. Implemented in the context of real options for mining by \cite{langrene15}, LOESS was enhanced through a ``sliding trick" that reduces complexity to $\mathcal{O}((N+N')\log N)$.

\subsection{Gaussian process regression (GPR)}
GPR is superficially similar to LOESS, in that the prediction $h(x_*)$ is a weighted average of sampled outputs $\by$. The underlying relationship $h: \emph{x} \mapsto y $ is taken to be a realization of a Gaussian random field, i.e.~$\{ h(x^i)\}_{i=1}^{N}$ is a sample from the multivariate Gaussian distribution with mean  $\{ m(x^i)\}_{i=1}^{N}$ and covariance function  $\{ \kappa(x^i,x^j)\}_{i,j=1}^{N}$.
Among many options in the literature, arguably the most popular is the squared exponential kernel,
$$\kappa(x^i,x^j) = \sigma_f^2 \exp \Big( -\frac{1}{2} (x^i - x^j)^T \mathbf{\Sigma}^{-1} (x^i - x^j) \Big), $$
where $\mathbf{\Sigma} $ is a diagonal matrix. Traditionally, diagonal elements  of the matrix $\mathbf{\Sigma}$  are known as the lengthscale parameters, and $\sigma_f^2$ as the signal variance. Together they are often referred to as hyper-parameters. While the lengthscale parameters determine the smoothness of the surface in the respective dimension, $\sigma_f^2$ determines the amplitude of the fluctuations. In figure \ref{fig:3dvaluefunction} we observed that for a fixed price, the continuation value function is linear in the inventory dimension, and has non-linear behavior in the price dimension. GPR captures this difference through its scale parameters, which will be ``large'' for inventory (slow decay of correlation, so little curvature) and ``small'' for price dimension (fast correlation decay allow for wiggles in terms of $P$). The hyperparameters are estimated through likelihood maximization. For the prior mean we take $m(x) = \beta_0$ where $\beta_0$ is learned together with the other hyperparameters. 

For any site $x_*$, $h(x_*)$ is a random variable whose conditional distribution given $\{ \bx,\by \}$ is:
\begin{equation}
    h(x_*)|\by \sim \mathcal{N}\Big(m(x_*)+ H_*\mathbf{H}^{-1}(\by-m(\bx)), H_{**}(x_*) - \mathbf{H}_{*}(x_*)\mathbf{H}^{-1}\mathbf{H}_{*}(x_*)^T \Big)
\end{equation}
where the $N \times N$ matrix covariance matrix $\mathbf{H}$ and the $N \times 1$ vector $\mathbf{H}_*(x_*)$ are
\begin{equation}
    \mathbf{H} := \begin{bmatrix}
    \kappa'(x^1,x^1) & \kappa(x^1,x^2) & \dots  & \kappa(x^1,x^N) \\
    \kappa(x^2,x^1) & \kappa'(x^2,x^2) & \dots  & \kappa(x^2,x^N)\\
    \vdots & \vdots & \ddots & \vdots \\
    \kappa(x^N,x^1) & \kappa(x^N,x^2) & \dots  & \kappa'(x^N,x^N)
\end{bmatrix}\!, \
    \mathbf{H}_*(x_*)^T := \begin{bmatrix}
    \kappa(x_*,x^1)\\
    \kappa(x_*,x^2)\\
    \vdots \\
    \kappa(x_*,x^N)
\end{bmatrix}, \
H_{**}(x_*) = \kappa'(x_*,x_*),
\end{equation}
where $\kappa'(x^i,x^j)=\kappa(x^i,x^j)+\sigma^2$.
Consequently, the prediction at $x_*$  is $\hat{h}(x_*)=m(x_*)+ \mathbf{H}_*(x_*)\mathbf{H}^{-1}(\by-m(\bx))$ and the posterior GP variance
$s^2(x_*) := H_{**}(x_*) - \mathbf{H}_{*}(x_*)\mathbf{H}^{-1}\mathbf{H}_{*}(x_*)^T$
provides a measure of uncertainty (akin to standard error) of this prediction. GPR generally has $\mathcal{O}(N^3+N N'^2)$ complexity, similar to kernel regression. GPR usually performs extremely well but becomes prohibitively expensive for $N \gg 1000$.

\begin{remark}
  There is an extensive GP ecosystem that allows for numerous extensions of the described GPR; particularly relevant aspects include more advanced prior mean specification for $m(\cdot)$; other kernel families $\kappa(\cdot, \cdot)$; heteroskedastic models that can handle state-dependent simulation noise $\sigma^2(\cdot)$; further techniques for selecting GP hyperparameters; piecewise models for allow for spatially non-stationary covariance. Since our focus is on expository presentation rather than searching for the best-performing approach (which is necessarily problem-dependent), we restrict attention to the basic GP methodology with off-the-shelf implementation.
\end{remark}

\section{Simulation Design}
\label{sec:exp design}

The second central piece of Algorithm \ref{algo_Generalized} concerns the design(s) $\CD_k$. Like the regression sub-problem, the template gives the user a wide latitude in selecting $\CD_k$ given a simulation budget $N$, with many feasible approaches. 
We identify four relevant aspects of potential designs: joint vs.~product; adaptive vs.~space-filling; deterministic vs.~stochastic; and unique vs.~replicated. Last but not least, we discuss the design size $N$.

To set the stage, let us summarize the ``conventional'' design \cite{balata17, boogert08, boogert13}. Traditionally, the design to solve the storage problem relied on a mix of global paths together with inventory discretization. In the price dimension, it consists of choosing $N^P$ initial conditions for the price  process $P_{t_0}$ at time $t_0$ and sampling $n=1,\ldots, N^P$ paths $P^n_\tii$ following the conditional density  $p(t_{k+1},.|t_k,P_{t_k})$, until terminal date $T$. The resulting collection $(P_\ti^{n})$ is used for the design ``mesh'' at each $\ti$. For the  endogenous inventory dimension, $I$ is discretized into $N^I$ levels: $I^l = l\Delta I$, $\Delta I = \frac{I_{\max} - I_{\min}}{N^I-1}$, $l=0,1,\ldots,N^I-1$. The overall design $\CD_{k}$ has $N=N^PN^I$ sites and is constructed as the Cartesian product $\{P_\ti^1,P_\ti^2,...,P_\ti^{N^P}\} \times \{I^0,I^1,...,I^{N^I-1} \}$. In our terminology, this is a product design that is adaptive and stochastic in $P$, space-filling and deterministic in $I$, and has no replication.  Its shape is convenient for the piecewise continuous regression scheme of Section~\ref{sec:piecewise} that treats the $P$ and $I$ coordinates separately. At the same time, such $\CD$ can be viewed as a cloud of sites in the joint $(P,I)$-domain and utilized with a bivariate regression scheme.

To summarize the range of simulation designs we use a short-hand nomenclature. Product designs are identified as $\CD_P \times \CD_I$, while joint designs are denoted with a single symbol. Subscripts are used where necessary to identify the dimensionality of the respective design. Different design types are identified by different letters. For example, we use $\cG$ for a grid design and $\cP$ for a density-based ``probabilistic'' design. The conventional design described above thus gets labelled as $\cP \times \cG$.

\subsection{Space Filling Designs}
To achieve the goal of learning $(P,I) \mapsto q(\ti, P,I, m)$ we need to explore continuation values throughout the input domain. A simple mechanism to achieve this is to spread out the design sites to fill the space. A gridded design, with design sites uniformly selected using a mesh size $\Delta$, already described for the conventional approach above, is an example of such \emph{space-filling} sequences.

Space -filling can be done either deterministically or randomly. For the deterministic case, besides the aforementioned grid $\cG$, one may employ various Quasi Monte Carlo (QMC) sequences, for example, the Sobol sequences $\cS$. Sobol sequences are useful in dimension $d > 1$ where they can produce a $d$-variate space filling design of any size $N$, whereas a grid is limited to rectangular constructions of size $N_P \times N_I$. QMC sequences are also theoretically guaranteed to provide a good ``uniform'' coverage of the specified rectangular domain. We experimented with the following two setups:
  \begin{itemize}
  \itemsep-0.3em
        \item Since we have only one exogenous variable $P$, we employ a 1-D Sobol sequence $\cS_1$ of size $N^P$ restricted to $[P_{\min},P_{\max}]$ at each time-step $t_k$, $k=0,\ldots,K$. We then discretize the inventory dimension as $\{I^1,I^2,...,I^{N^I} \}$, similar to conventional design, and the final $\CD$ is the product $\cS_1 \times \cG $.
        \item Alternatively, we generate $N$ design sites from the two-dimensional Sobol sequence $\cS_2 = \{ P^n,I^n \}_{n=1}^{N} $ on the restricted domain $[P_{\min},P_{\max}] \times [I_{\min},I_{\max}]$.
    \end{itemize}

 An example of a randomized space filling design is taking $P^n_\ti \sim Unif(P_{\min},P_{\max})$ i.i.d. Because uniform samples tend to cluster, there are variance-reduced versions, such as Latin hypercube sampling (LHS). In two dimensions, the LHS design $\cL_2$ divides the input space into a rectangular array and ensures that each row and column has exactly one design site.

Note that if the same deterministic space-filling method is employed across time-steps, the design $\CD_\ti \equiv \CD$ becomes identical in $t_k$. This may generate ``aliasing'' effects from the regression scheme, i.e.~approximation artifacts around $(P^n, I^n)$ due to the repeated regressions and respective error back-propagation. Changing or randomizing $\CD_\ti$ across $\ti$'s is one remedy and often preferred as an implementation default. Another issue with space-filling designs is the need to specify the bounding box $[P_{\min}, P_{\max}] \times [I_{\min}, I_{\max}]$--which is easy in $I$ but not obvious in the $P$-dimension where $P_t$ is unbounded.

\subsection{Adaptive Designs}

 In contrast to space-filling designs that aim to cover the input space (and statistically target the global $L^2(Leb)$  approximation error), adaptive designs try to exploit the specific structure of the problem at hand. Their starting point is the observation that the quality of $\hat{V}_0$ depends on the correct prediction of storage actions along \emph{controlled} paths $(P^{n'}_\ti, \hat{I}^{n'}_\ti)$. Information about the distribution of the system state  was already leveraged in the conventional approach which used randomized, \emph{probabilistic} design for $\CD_P$. Similarly, \cite{boogert13} used a non-uniform discretization in $I$ to put more mesh points closer to $I_{\min}$ and $I_{\max}$.

 The above discussion motivates more general adaptive designs that aim to match the shape of $\CD$ to a proposal density $\mathfrak{p}(\cdot)$ (either bivariate or univariate for $P$ and $I$). For instance, a randomized joint probabilistic design $\cP_2$ samples from the bivariate distribution of $(P, \hat{I})$. An alternative is to use the respective marginal densities to build a probabilistic product design $\cP \times \cP$, taking $P^n_{\ti} \sim \mathfrak{p}^P(t_k,\cdot)$ and inventory $I^n_{\ti} \sim \mathfrak{p}^I(t_k,\cdot)$. One may also build deterministic adaptive designs by quantizing the proposal densities $\mathfrak{p}$ (similar to numerical quadrature methods) to return a discrete representation with $N^P$ sites.    Probabilistic designs ensure better estimation of the continuation function in the region of the domain where the $(P_{0:T}^{n'},\hat{I}_{0:T}^{n'})$ trajectory is most likely to be.

 \begin{remark}
 To construct the distribution $\mathfrak{p}^I(\ti_,\cdot)$  of the endogenous variable $\hat{I}_\ti$ requires some foresight. One strategy is to use a small part of the simulation budget to create a policy using traditional/space-filling design and run a forward simulation on this policy, saving the result of the forward simulations as a proxy for joint distribution in creating design for optimal control map.  
 \end{remark}

\subsection{Batched Designs}

Non-parametric techniques like GPR and LOESS generally bring improved accuracy at the cost of increased regression overhead. Indeed, for both methods the complexity is at least quadratic in the number of sites $N$ which can become prohibitive for $N \gg 1000$. One solution to overcome this hurdle is to use replicates, i.e.~re-use the same design site for multiple simulations. Thus, rather than having thousands of distinct design sites equivalent to  the simulation budget $N$, we select only a few hundred distinct sites $N_s$ and generate $N_b:=N/N_s$ paths from each design site. Formally this means that we distinguish between the $N$ initial conditions $(P^n_\ti, I^n_\ti)_{n=1}^N$ for simulating pathwise continuation values and the \emph{unique} design sites $N_s \ll N$ which comprise the design $\bar{\CD}$. The latter can then be of any type, space-filling, adaptive, etc.

Given a design site $(P^n_\ti,I^n_\tii)_{n=1}^{N_s}$, we make $N_b$ draws $P^{n,(m)}_\tii$ and evaluate the corresponding pathwise continuation value $v_{{k+1}}^{(m)}(P^{n,(m)}_\tii,I^n_\tii)$, $m=1,\ldots, N_b$. For kernel-based techniques like GP and LOESS one may then work with pre-averaged values, i.e.~first  evaluate the empirical average :
    $$\bar{v}^n_{{k+1}}(P^n_\ti,I^n_\tii) = \frac{1}{N_b} \sum_{m=1}^{N_b} v_{{k+1}}^{(m)},$$
    across the $N_b$ replicates. One then feeds the resulting regression dataset $(P^n_\ti,I^n_\tii, \bar{v}^n_{{k+1}} )_{n=1}^{N_s}$ into the regression equations to estimate the continuation function.

    Besides reducing the overall time spent in regressions (which can easily be several orders of magnitude), a batched design has an advantage of reduced simulation variance of $\bar{v}$ at each design site, thus improving the signal-to-noise ratio. While a replicated design is sub-optimal (in terms of maximizing the quality of the statistical approximation), in practice for large $N$ (which are frequently in the hundreds of thousands) the loss of fidelity is minor and is more than warranted given the substantial computational gains.

    Other than uniform batching approach that uses a fixed $N_b$ number of replicates at each site, one could also employ adaptive batching (i.e.~$N_b$ as site-specific) with more replications around the switching boundaries to gain better precision \cite{ludkovski16}.

\subsection{Dynamic Designs}
DEA lets us easily combine different designs at various time steps. For example, one may  vary the step-wise simulation budget, employing larger $N$ near maturity to effectively capture the effect of the terminal conditions in the continuation function, and lesser budget thereafter. Some regression methods may be computationally expensive but more ``accurate" (GPR), while others may be less ``accurate" but computationally efficient (polynomial regression). Judicious combination of the two can save time without much loss in fidelity.

\begin{figure*}[!ht]
        \centering
        \begin{subfigure}[b]{0.375\textwidth}
            \centering
            \includegraphics[width=\textwidth]{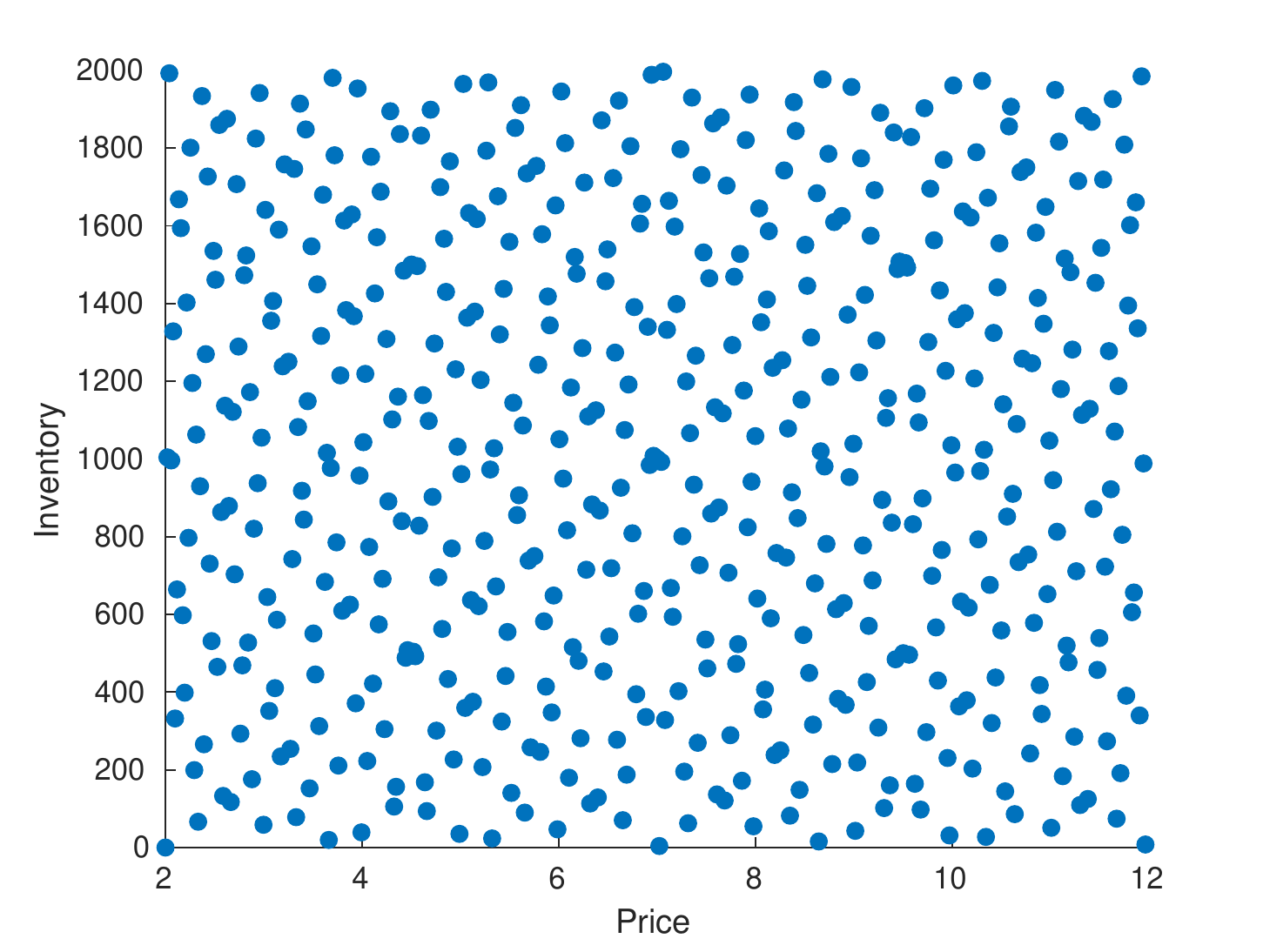}
            \caption[]%
            {{\small 2D Sobol QMC sequence $\cS_2$}}
            \label{fig:sobolDesign}
        \end{subfigure}
        \quad
        \begin{subfigure}[b]{0.375\textwidth}
            \centering
            \includegraphics[width=\textwidth]{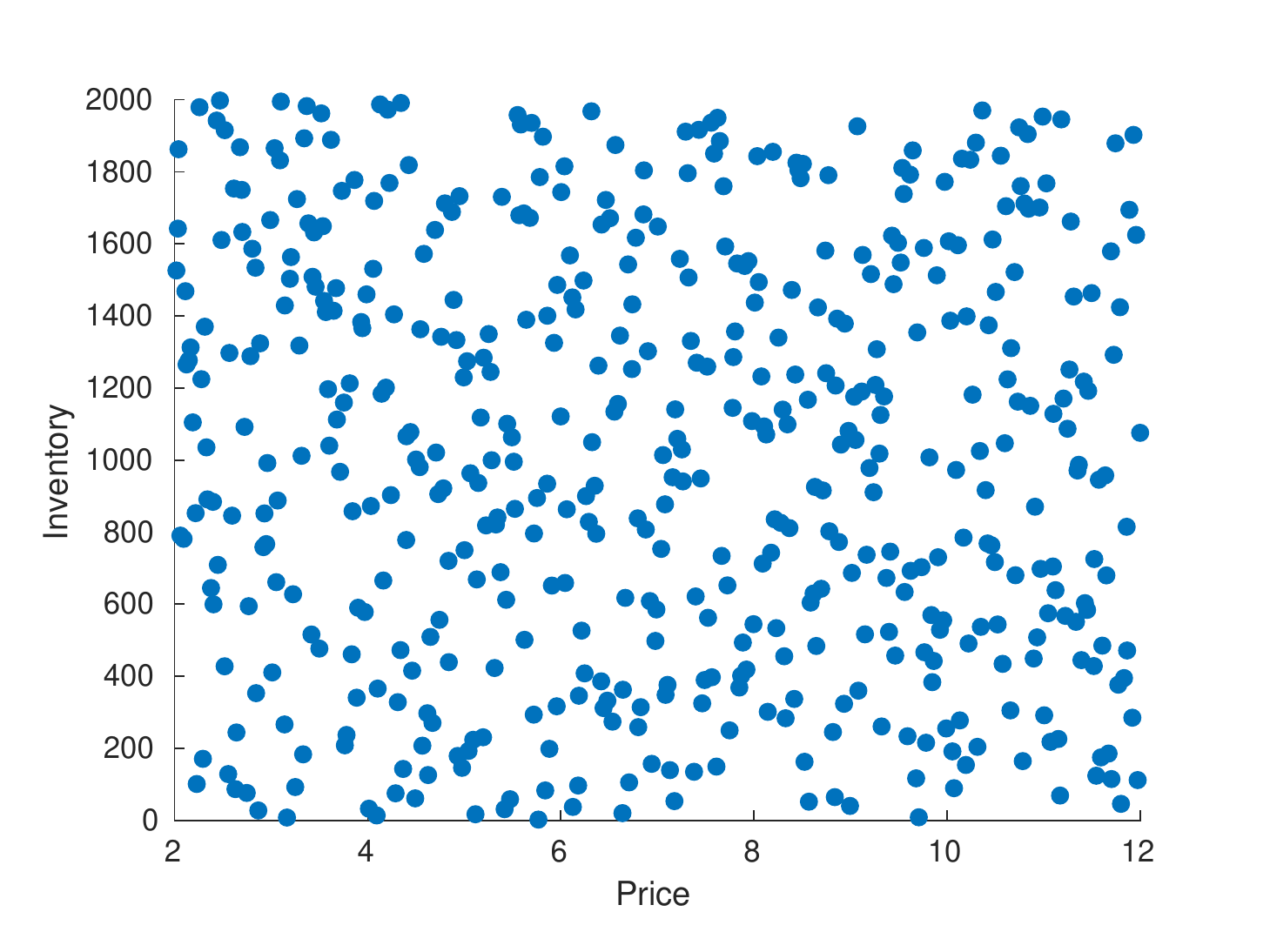}
            \caption[]%
            {{\small 2D LHS design $\cL_2$}}
        \end{subfigure}
        \begin{subfigure}[b]{0.375\textwidth}
            \centering
            \includegraphics[width=\textwidth]{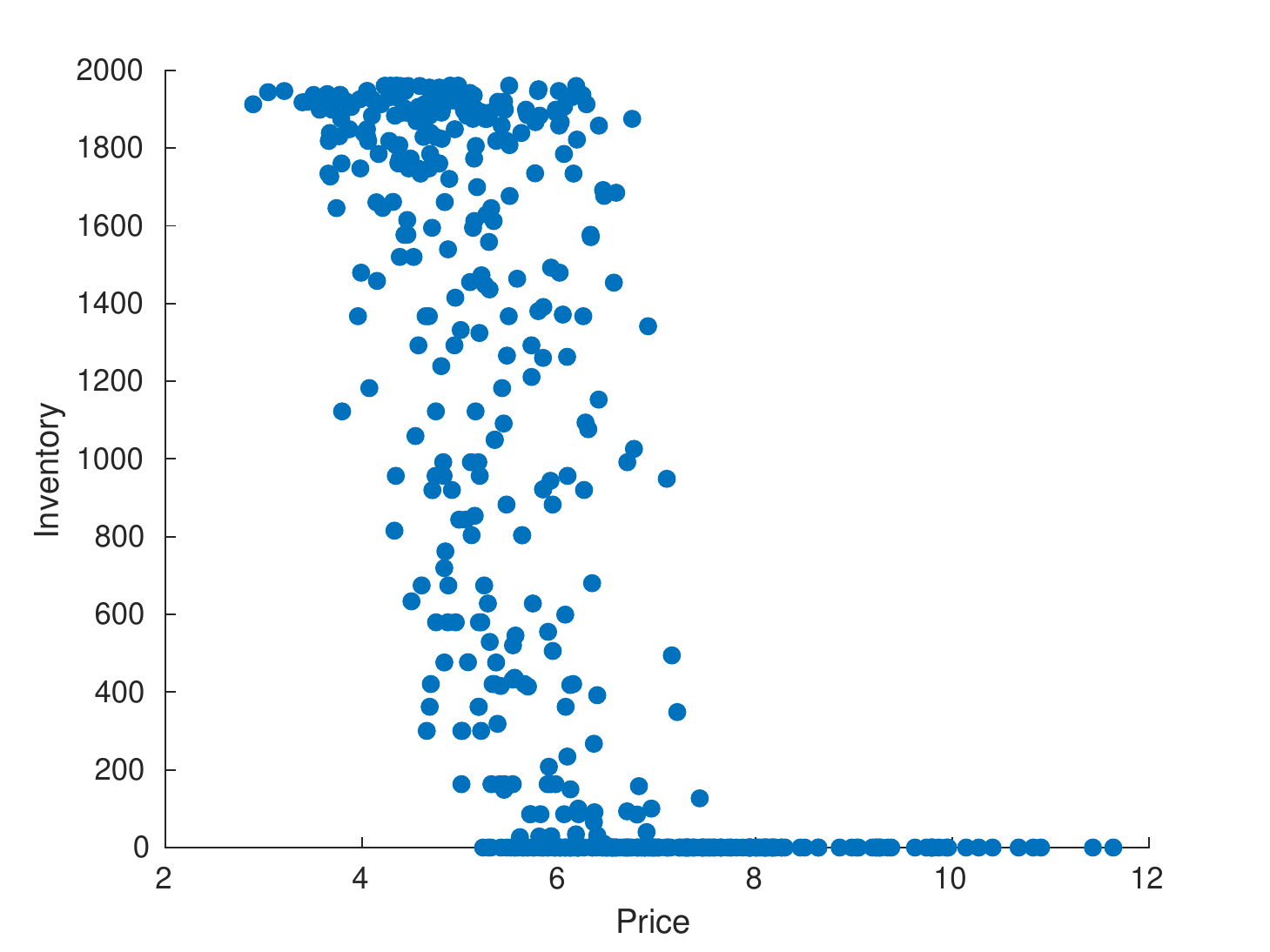}
            \caption[]
            {{\small Joint Probabilistic design $\cP_2$ }}
            \label{fig:adaptiveDesign}
        \end{subfigure}
        \quad
        \begin{subfigure}[b]{0.375\textwidth}
            \centering
            \includegraphics[width=\textwidth]{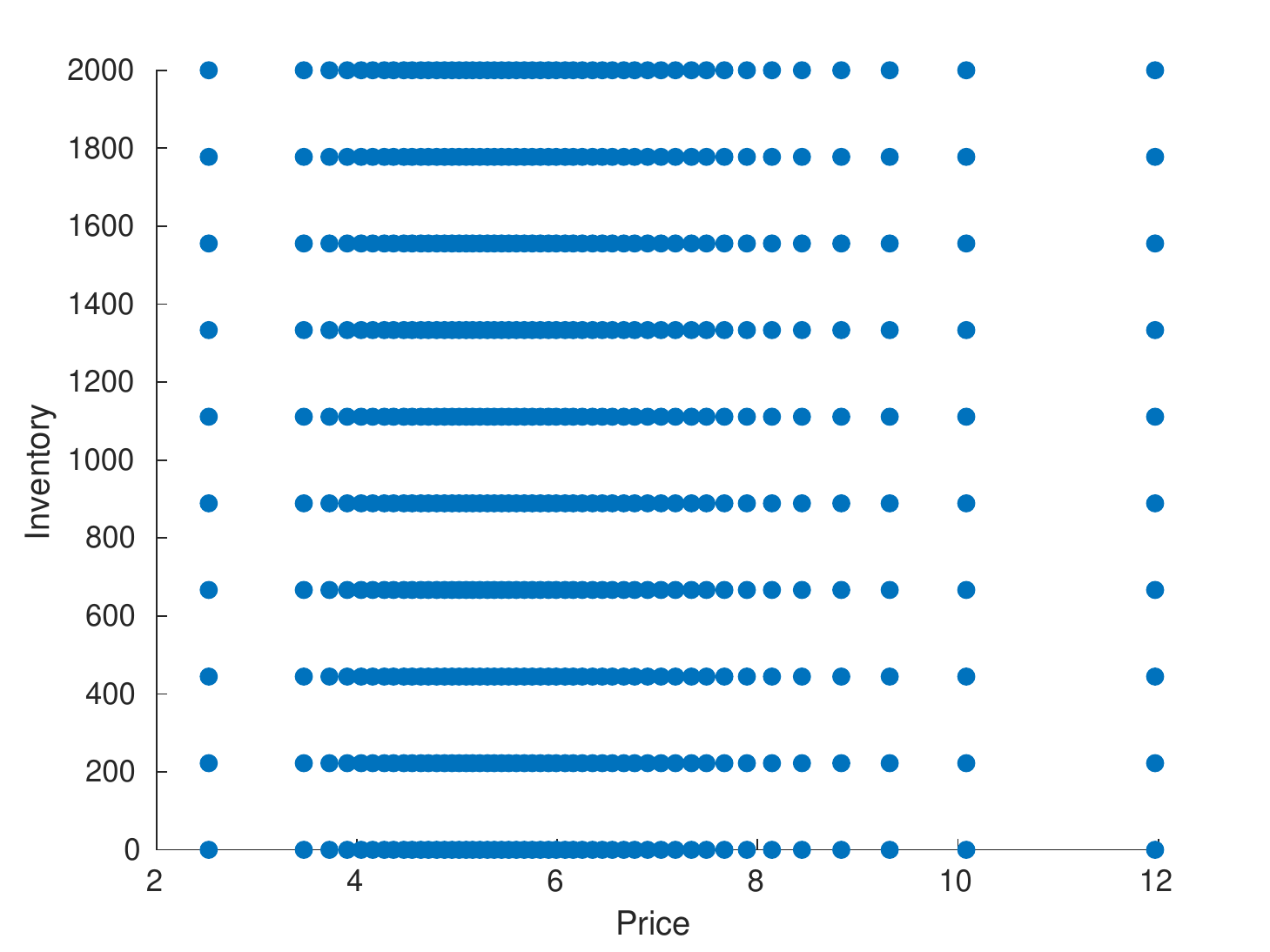}
            \caption[]%
            {{\small Conventional product design $\cP \times \cG$}}
        \end{subfigure}

        \caption[]
        {\small {Illustration of different simulation designs $\CD$. In all cases we take $N=500$ design sites. }}
        \label{fig:designs}
    \end{figure*}

We conclude this section with two illustrations. In Figure~\ref{fig:designs} we display four representative designs: Sobol space-filling sequence in 2D $\cS_2$, LHS in 2D $\cL_2$, joint probabilistic $\cP_2$, and conventional design $\cP \times \cG$. These will also be used in the numerical examples below. While Sobol sequence fills the input space with a symmetric pattern, LHS is randomized. The joint probabilistic design mimics the distribution of the state variables $(P_t,\hat{I}_t)$, putting most sites at the boundaries of the inventory  $I_{min}, I_{max}$ and around the mean of the price. Note that this design is very ``aggressive'' and does not explore enough of the input space which can make the regression estimates be non-robust. Therefore, in the following numerical examples we blend (via a statistical mixture) probabilistic designs with other types. Finally, conventional design discretizes the inventory while maintaining the adaptive distribution in the price dimension.

\begin{figure*}[!ht]
    \centering
    \includegraphics[scale=0.35, trim=0in 0.5in 0in 0.5in]{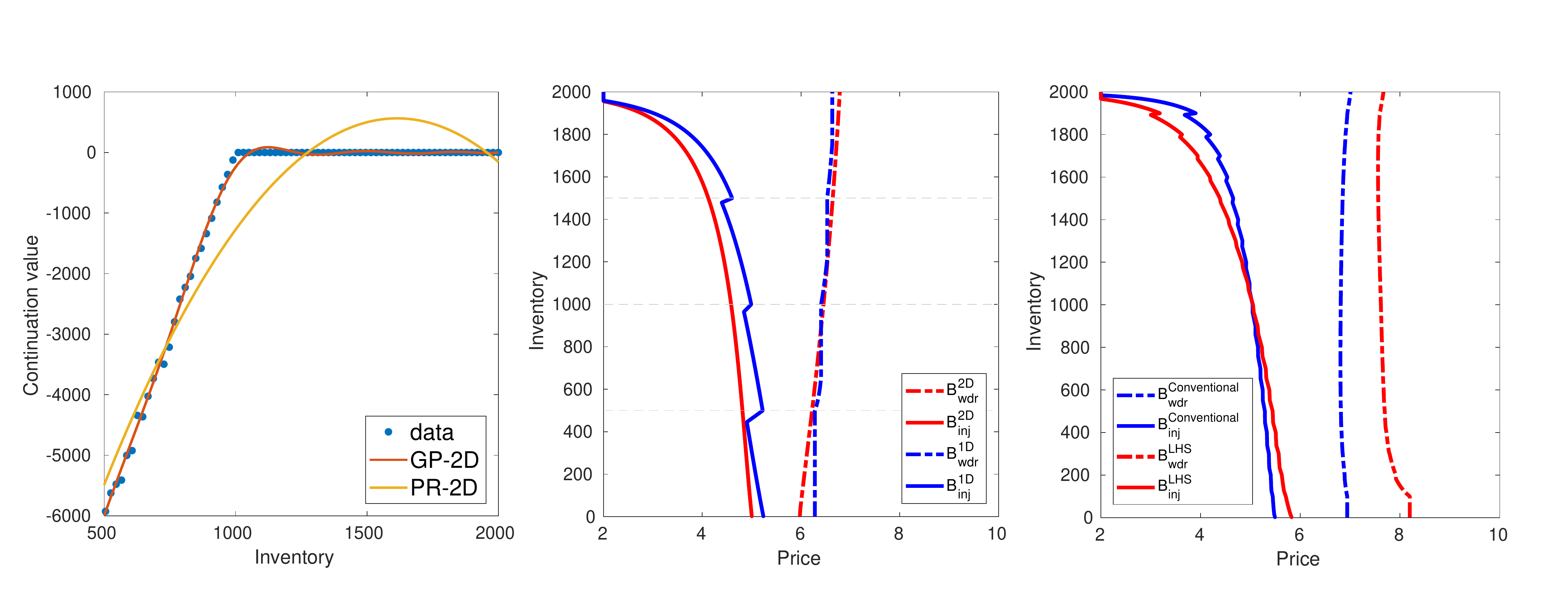}
    \caption{\emph{Left panel:} Comparison of the continuation function  $q(t,P, \cdot)$ for different regressions at one step to maturity $t = T - \Delta t$ and $P=6$. \emph{Center panel:} the control maps $\hat{m}(t,P,I)$ at $t=2.7$ years corresponding to GP-1D and GP-2D regressions. For GP-1D we used $N_s=100$, $N_b=20$ and $N_I=5$. For GP-2D we used $N_s=500$ and $N_b=20$. The horizontal lines represent the inventory  discretization levels for the GP-1D design. \emph{Right panel:} the control maps corresponding to conventional and LHS design with PR-1D regressions. We used $N^P=2000$, $N^I=21$ and price domain for LHS $[P_{min},P_{max}] = [2,20]$. }
    \label{fig:regressionDesignEffect}
\end{figure*}

Next, in Figure \ref{fig:regressionDesignEffect} we display the effect of regression and design on the continuation value function and the corresponding control maps. The left panel compares the continuation value function of GP-2D and PR-2D at the last step before maturity $t=T-\Delta t$. We observe that PR-2D has a poor fit compared to GP-2D, which almost perfectly match the ``hockey-stick'' penalty. Furthermore, -1D regressions have ``non-smooth'' switching boundaries due to the piecewise regressions in $I$. This is evident in the center panel of the figure, where the control map shows jumps at $I \in \{500,1000,1500 \}$, the intermediate discretization levels of inventory (we used $N^I = 5$ with $\Delta I = 500$). This behaviour becomes less prominent when $N^I$ is increased. The right panel of the figure visualizes two control maps for PR-1D with LHS and conventional design. The effect of conservative domain (we used $[P_{min},P_{max}] = [2,20]$) for LHS is evident as we notice that the Store region is much too wide in the LHS variant, not allowing the controller to benefit from withdrawing when prices are $P \in [7, 7.5]$ resulting in lower valuation.

\section{Natural Gas Storage Facility}
\label{sec:example1}

\subsection{Market and Storage Description}\label{sec:storage-1}
In this section we revisit the gas storage problem in the setting of Chen and Forsyth~\cite{forsyth07}. The exogenous state process $P$  follows logarithmic mean-reverting dynamics
 \begin{equation}
P_\tii - P_\ti = \alpha (\underline{P} -P_\ti)\Delta t + \sigma P_\ti \, \Delta W_\ti,
\label{eqn:priceProcess}
\end{equation}
where $P_t$ is the spot price per unit of natural gas and is quoted in ``dollars per million of British thermal unit (\$/MMBtu)". The inventory $I_t$ is quoted in million cubic feet (MMCf). Since roughly 1000MMBtu = 1MMcf, we multiply $P_t$ by $10^3$ when calculating revenue or profit.

The penalty function $W(P_T,I_T)$ at maturity is
$W(P_T,I_T) = -2P_T\max(1000-I_T,0).$
Thus, the target inventory is $I_T = 1000$ or 50\% capacity. There is no compensation for excess inventory and a strong penalty (at 200\% of the market price) for being short. As a result, the value function at maturity has a non-smooth hockey-stick shape in $I$, with zero slope for $I_T < 1000$ and a slope of $-2P_T$ otherwise.

The withdrawal and injection rates are inventory-dependent and given by
$$c_{wdr}(I) = -k_1\sqrt{I} \ \text{ and } \ c_{inj}(I) = k_2 \sqrt{\frac{1}{I +k_3} - \frac{1}{k_4}}.$$
These functional specifications are derived from the physical hydrodynamics of the gas storage facility, see \cite{rasmussen09}. The resulting dynamics of the inventory is:
\begin{equation}
    I_\tii = I_\ti +
    \begin{cases}
      c_{wdr}(I_\ti) \Delta t, & \text{if } m_\ti = -1 \quad\text{(withdrawal)}; \\
      0  & \text{if } m_\ti = 0; \\
      (c_{inj}(I_\ti) - k_5)\Delta t, & \text{if } m_\ti = +1 \quad  \text{(injection).}
    \end{cases}
\label{storageConst2}
\end{equation}
The constant $k_5$ measures the cost of injection which is represented as ``gas lost''. It leads to a profit gap  between production and injection whereby no-action $(m^*=0)$ will be the optimal action if the price is ``close" to the mean level, $P_t \approx \underline{P}$.

In the numerical experiments below we discretize $T$ into $K=1000$ steps, so that $\Delta t = 0.001T$, the rest of the parameters are listed in Table  \ref{storageConstParam}. The switching costs are taken to be zero $K({i,j}) \equiv 0 \forall i,j$. Absence of switching cost reduces the state variables in the continuation function  $q(\ti, \cdot)$ in~\eqref{eqn:RNMC}  to $(P_\ti,I_\tii)$. As a result, in this example every time step requires one projection of the 2D value function.
\begin{table}[!ht]
\centering
\begin{tabular}{c} \hline 
$\alpha = 2.38, \ \sigma = 0.59, \ \underline{P} = 6$ \\ \hline
$k_1 = 2040.41, \ k_2 = 7.3 \cdot 10^5,\ k_3 = 500,\ k_4 = 2500,\ k_5 = 620.5$ \\ \hline
$I_{\max} = 2000$ MMcf,\ $T=3$,\ $\Delta t = 0.003$,\ $r=10\%$ \\
\hline
\end{tabular}
\caption{Parameters for the gas storage facility in Section~\ref{sec:example1}.}
\label{storageConstParam}
\end{table}

\subsection{Benchmarking Setup}

 To benchmark the performance, we compare to two schemes utilizing conventional product design $\cP \times \cG$ (with inventory uniformly discretized): degree-3 global polynomial approximation over $(P,I)$ (PR-2D) and piecewise continuous approximation (with degree-3 polynomial regressions in $P$ at each inventory level $I_k$, PR-1D). These can be viewed as a ``classical'' TvR scheme with a joint regression \cite{ludkovski10,denault2013}, and the discretized-$I$ version as used by  \cite{balata18,balata17,boogert08,boogert13}. Additionally, we implement five regression approaches (PR-1D, GP-1D, PR-2D, LOESS-2D and GP-2D) on several different designs, with simulation budgets: $N \simeq 10K, 40K, 100K$:

\begin{itemize}

\item  Randomized space filling design implemented via LHS ($\cL_1 \times \cG$ for piecewise-continuous regression and $\cL_2$ otherwise). We use a large conservative input domain $P \in [2,10]$.

\item  Mixture-2D design with 40\% of sites from space-filling and the remaining 60\% from the joint empirical distribution $\cP_2$, $\CD_M := \cP_2(0.6N) \cup \cL_2(0.4N)$.  To implement $\cP_2$, we need to estimate $\mathfrak{p}^P(t_k,\cdot)$ and $\mathfrak{p}^I(t_k,\cdot)$. This is done offline by first running the algorithm with a small budget and conventional product design. We then generate forward paths $(P^{n'}_\ti, \hat{I}^{n'}_{\ti})$  to estimate the joint  $(\mathfrak{p}^P(t_k,\cdot), \hat{\mathfrak{p}}^I(t_k,\cdot))$ at $\ti$. Since the  marginal distribution $\hat{\mathfrak{p}}^I(t_k,\cdot)$ starts to  concentrate around $I=1000$ as we get close to the maturity of the contract, the resolution at other parts of the domain is reduced and $\cL_2$ is to used to compensate for this effect.

\item Adaptive-1D: For 1D regressions, we estimate $\mathfrak{p}^P(t_k,\cdot)$ as above, and then non-uniformly discretize the inventory to incorporate the fact that the optimally controlled inventory process $\hat{I}_\ti$ concentrates around $I_{\min},I_{\max}$ and $I=1000$. Discretization levels for each simulation budget are detailed in Appendix~\ref{app}.

\item Dynamic time-dependent design that varies the simulation budget $N_\ti$ across $\ti$. We used the specification $N_{t_k}(N_{(1)}, N_{(2)}) := N_{(1)}\mathbf{1}_{\{k < 900\}} + N_{(2)} \mathbf{1}_{\{k \ge 900\}}$ 
such that (approximately) $0.9 N_{(1)} + 0.1 N_{(2)} \in \{ 10K, 40K, 100K \}$. Exact specification is given in Table~\ref{table:designSpecification} in the Appendix. We use two variants of it:
\begin{itemize}
\itemsep-0.3em
\item fixed projection space $\cH$ (namely GP-1D) and conventional product design $\CD = \cP \times \cG$.
\item time-dependent projection and designs, namely GP-2D and Mixture design for $k < 900$ and PR-1D and Conventional design for $k \ge 900$. 
\end{itemize}
The motivation for the above Dynamic scheme is to better handle the non-smooth terminal condition by devoting to it larger simulation budget, as well as using the -1D regression.
\end{itemize}

For GP we use \texttt{Matlab}'s in-built implementation \texttt{fitrgp}. For LOESS, we use the \texttt{curvefitting} toolbox again from \texttt{Matlab} 
that constructs local quadratic approximations based on the tri-cube weight function $\kappa(x_*,\emph{x}^n) = \Big(1 - (\frac{|x_* - x^n|}{\lambda(d,x_*)})^3\Big)^3$. Above $\lambda(d,x_*)$ is the Euclidean distance from $x_*$ to the most distant $x^k$ within the span $d$. We use the default span of $d=25\%$, keeping $P$ on its original scale and re-scaling $I$ to be in the range $[0,2]$. 
To reduce the regression overhead of both GPR and LOESS we utilize batched designs with $N_b$ replicates (see Appendix \ref{app}) on top of the underlying design type.

In order to compare the performance of different designs/regressions, we use the estimate of the value function $\hat{V}(0,P_0,I_0)$ at $P_0=6, I_0=1000$ using a fixed set of $N'=10,000$ out-of-sample paths (i.e.~fixed $P^{n'}_{0:T}$) as a performance measure.

\subsection{Results}
\label{sec:results_example1}

Table~\ref{table: DOE} presents the performance of different designs and regression methods. We proceed to discuss the results focusing on three different aspects: (i) impact of different regression schemes, in particular parametric PR vs.~non-parametric GP and LOESS approaches; (ii) impact of simulation design; (iii) joint -2D vs.~interpolated -1D methods.

First, our results confirm that the interpolated -1D method performs extremely well in this classical example, perfectly exploiting the 2-dimensional setup with a 1-dimensional inventory variable. In that sense the existing state-of-the-art is already excellent. There are two important reasons for this. First  PR-1D is highly flexible with lots degrees of freedom, allowing a good fit (with overfitting danger minimized due to a 1D setting). Second, PR-1D perfectly exploits the fact that the value function is almost (piecewise) linear in inventory. We obtain a slight improvement by replacing PR-1D with GP-1D; another slight gain is picked up by replacing the equi-spaced inventory discretization with an adaptive approach that puts more levels close to the inventory boundaries. As a further enhancement, the Dynamic design utilizes a step-dependent simulation budget (to capture the boundary layer effect due to the non-smooth ``hockey-stick" terminal condition that requires more effort to learn statistically), leading to significant improvement, highlighting the potential benefit of mixing-and-matching approximation strategies across time-steps. By taking $N$ time-dependent one may effectively save simulation budget (e.g.~the valuation for Dynamic GP-1D with $N=10^4$ is comparable to $N=2 \cdot 10^4$ for Adaptive GP-1D). Nevertheless, as we repeatedly emphasize the -1D methods are necessarily limited in their scope, in particular not scaling as more factors/inventory variables are added.  A further limitation of the -1D method is their requirement of a product design.

The much more generic bivariate -2D regressions give a statistically equitable treatment to all state variables and hence permit arbitrary simulation designs. The resulting huge scope for potential implementations is both a blessing and a curse. Thus, we document both some good and some bad choices in terms of picking a regression scheme, and picking a simulation design. We find that PR-2D tends to significantly underperform which is not surprising given that it enforces a strict parametric shape for the continuation value with insufficient room for flexibility. Similarly, we observe middling performance by LOESS; on the other hand GPR generally works very well.

Turning our attention to different 2D simulation designs, we compare the conventional $\mathcal{P} \times \mathcal{G}$ choice against 3 alternatives: conservative space-filling $\mathcal{L}_2$ on a large input domain; joint probabilistic design $\mathcal{P}_2$; a mixture design that blends the former two. We find that both plain space-filling and joint probabilistic do not work well; the first one is not targeted enough, spending too much budget on regions that make little contribution to $\hat{V}$; the second is too aggressive and often requires extrapolation produces inaccurate predictions when computing $\hat{m}$. In contrast, the mixture design (we found that a 60/40 mix works well) is a winner, significantly improving upon the conventional one. In particular, GP-2D with Mixture design is the only bivariate regression scheme which performs neck-to-neck with GP-1D. This is significant because unlike GP-1D, GP-2D can be extended to higher dimensions in a straightforward manner and does not require a product design (or any interpolation which is generally slow). These findings highlight the importance of proposal density in design choice. To highlight the flexibility of our algorithm, we also present another dynamic design combining Adaptive PR-1D with Mixture GP-2D. This combination maintains the same accuracy but runs about 10-15\% faster thanks to lower regression overhead of PR-1D.

\begin{table}[!ht]
\centering
\begin{tabular}{|l|l|r r r|}
\hline
 & Regression & \multicolumn{3}{c|}{Simulation Budget} \\
Design & Scheme & Low & Medium & Large \\ \hline \hline
\multirow{4}{*}{Conventional} & PR-1D & 4,965 & 5,097 & 5,231  \\
& GP-1D & 4,968 & 5,107 & 5,247 \\ \cline{2-5}
 & PR-2D & 4,869 & 4,888 & 4,891  \\
 & LOESS-2D & 4,910 & 4,969 & 5,011  \\
 & GP-2D & 4,652  & 5,161  &  5,243 \\ \hline \hline
\multirow{5}{*}{Space-filling} & PR-1D & 4,768 & 4,889 & 5,028  \\ %
& GP-1D & 4,854 & 5,064 & 5,224  \\ \cline{2-5}
& PR-2D & 4,762 & 4,789 & 4,792  \\ 
 & LOESS-2D & 4,747  & 4,912  & 4,934  \\
 & GP-2D  & 4,976 & 5,080 & 5,133 \\ \hline \hline
\multirow{2}{*}{Adaptive 1D} & PR-1D & 5,061 & 5,187 & 5,246 \\ 
& GP-1D & 5,079 & 5,195 & 5,245 \\ \hline 
\multirow{2}{*}{Dynamic} & GP-1D & 5,132 & 5,225 & 5,266 \\
& Mixed & 5,137 & 5,205 & 5,228 \\  \hline
\multirow{3}{*}{Mixture 2D} & PR-2D & 4,820 & 4,835 & 4,834 \\ 
& LOESS-2D & 4,960 & 4,987 & 5,003  \\
& GP-2D  & 5,137 & 5,210 & 5,233  \\  \hline
\end{tabular}
\caption{Valuation $\hat{V}(0,6,1000)$ (in thousands) using different design-regression pairs and three simulation budgets: Low $N\simeq 10K$, Medium $N\simeq 40K$, Large $N\simeq 100K$, cf.~Appendix \ref{app}. The valuations are averages across 10 runs of each scheme except for LOESS-2D with large budget: due to the excessive overhead of LOESS only a single run was carried out.}
\label{table: DOE}
\end{table}

\textbf{Impact of $P$-Designs for -1D Methods.}
In the context of -1D methods that focus on regressing in $P$ and discretizing + interpolating in $I$, changing the design from $\cP \times \cG$ to a space-filling $\cL \times \cG$  has an adverse effect on all the methods, for example, PR-1D observes a drop of valuation by over 200K (cf. right panel, Figure \ref{fig:regressionDesignEffect}). Moreover, intermediate designs which use space-filling but ``crudely'' mimics the distribution of $P_t$ (for example, 60\% of the sites in the range $P \in [4.5,7.5]$, 20\% below $P=4.5$ and 20\% above $P=7.5$) interpolate the valuation between $\cL \times \cG$  and  $\cP \times \cG$, confirming the significant effect of design in $P$ dimension.

The shape of the simulation design in the $I$-dimension has an equally important role in the valuation. This can be observed by comparing the performance of Conventional PR-1D and Adaptive PR-1D (Table~\ref{table: DOE}) that share the same $\mathcal{P}$ P-design based on the marginal distribution of $P_t$ but utilize different approaches to discretize the inventory. 
The non-uniform discretization in the Adaptive design improves precision close to $I_{\min}$, $I_{\max}$ and $I=1000$ and leads to a higher valuation relative to Conventional, with an additional advantage that we do not need to store hundreds of thousands of paths in memory.

\begin{figure*}[!ht]
    \centering
    \includegraphics[height=1.9in, trim=0.5in 0.5in 0.5in 0.5in]{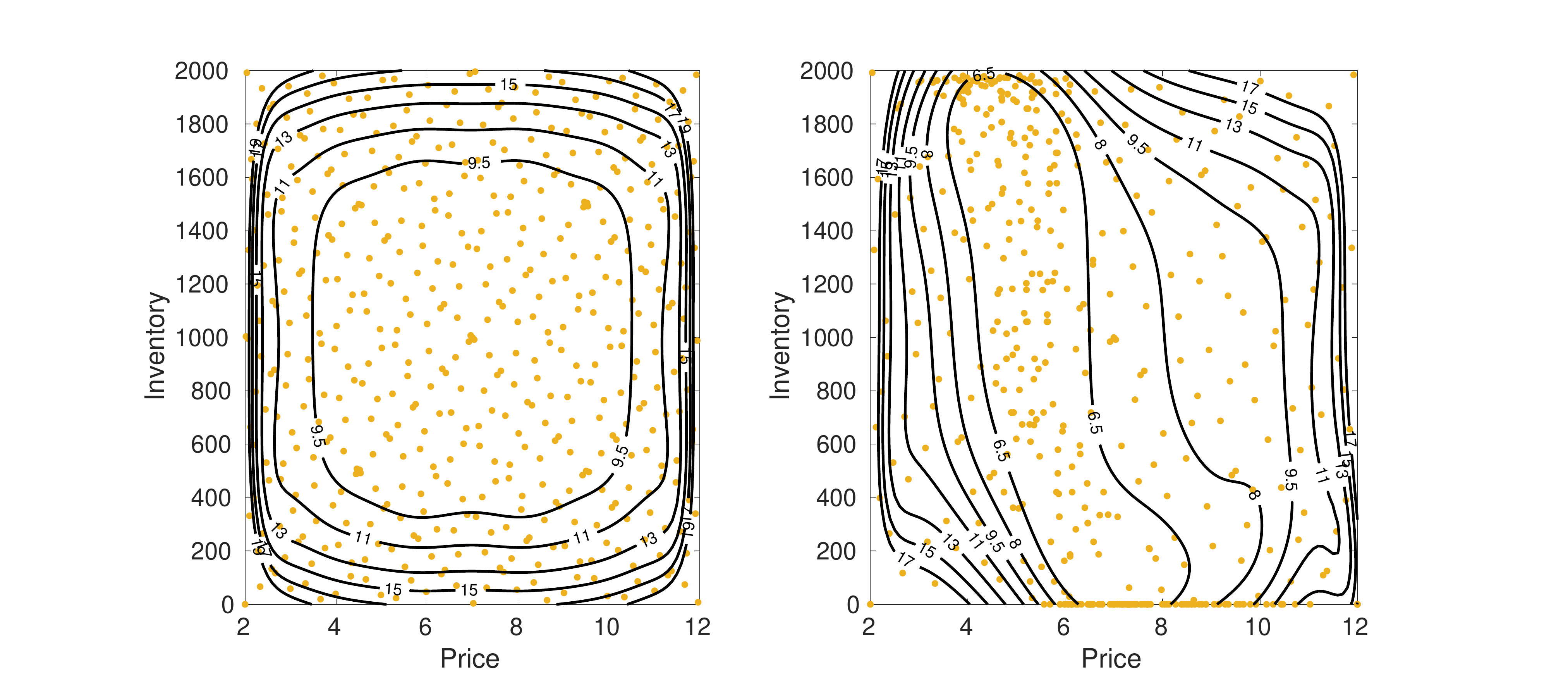}
    \caption{ Impact of design on the posterior standard error $s(x)$ of GPR at $t=1.5$ years. We show the contours of $x \mapsto s(x)$ with dots indicating the underlying respective designs with $N=500$ sites.  
    \emph{Left panel:} Space-filling design (Sobol QMC $\cS_2$) of Figure~\ref{fig:sobolDesign} leads to rectangular level sets of $s(\cdot)$. \emph{Right panel:} under a mixed design $s(\cdot)$ resembles the shape of $\cP_2$ in Figure~\ref{fig:adaptiveDesign}. }
    \label{fig:contourPlot}
\end{figure*}

\textbf{Visualizing Design Impact with GPR.}
Besides higher valuation, GPR also provides a natural way to quantify the effect of design through posterior standard error. In Figure~\ref{fig:contourPlot} we plot the posterior GP standard deviation $s(x)$ (proxy for local estimation standard error) based on Mixture and Space-filling 2D designs. This visualizes the direct link between the design shape and GPR learning precision. For a space-filling design we observe a constant posterior variance, i.e.~GPR learns $\hat{q}(\ti, \cdot)$ equally well across the interior of the regression domain. In contrast, the posterior standard deviation is very high around the edges of $[P_{\min}, P_{\max}] \times [I_{\min}, I_{\max}]$, which is problematic for correctly identifying the strategy when inventory is almost full or almost empty. For the Mixture design, the posterior variance reflects the concentration of the input sites along the diagonal, compare to the joint probabilistic design $\cP_2$ in Figure~\ref{fig:adaptiveDesign}. In turn, this is beneficial for learning the control map, as the GPR prediction is most accurate along the switching boundaries, cf.~Figure~\ref{fig:controlMapEvolution_b}. Higher precision around the switching boundaries allows the Mixture design to allocate the simulation budget to the regions where accuracy is needed most and partly explains its higher valuations compared to an LHS design. We emphasize that such local inference quality is only available with non-parametric tools; global schemes like PR cannot directly benefit from focused designs.

\begin{figure}[!ht]
  \centering
  \begin{subfigure}[b]{0.45\textwidth}
    \includegraphics[width=\textwidth]{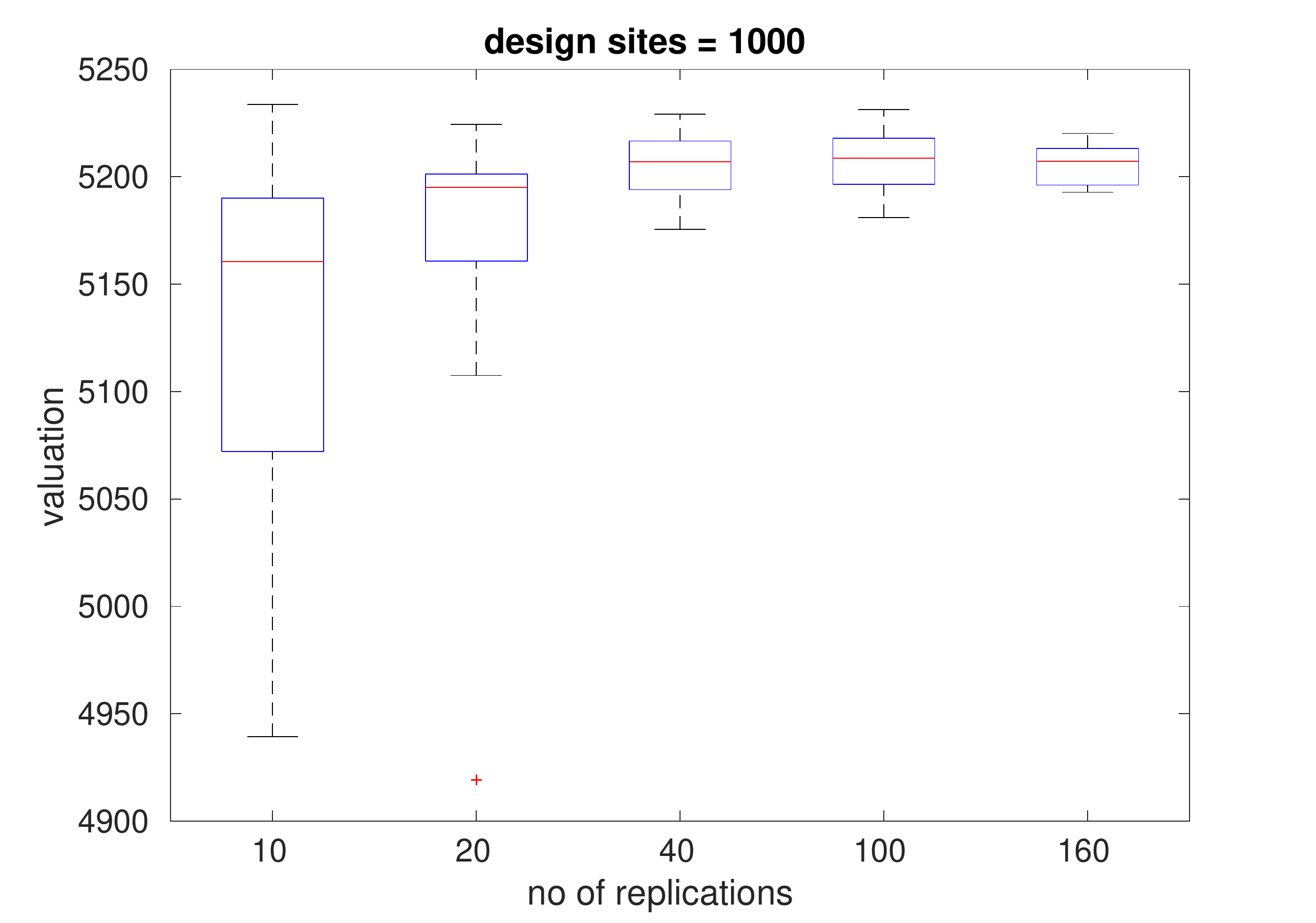}
    \caption{Impact of replication $N_b$}
    \label{fig:replicationEffect}
  \end{subfigure}
 \quad
    \begin{subfigure}[b]{0.45\textwidth}
    \includegraphics[width=\textwidth]{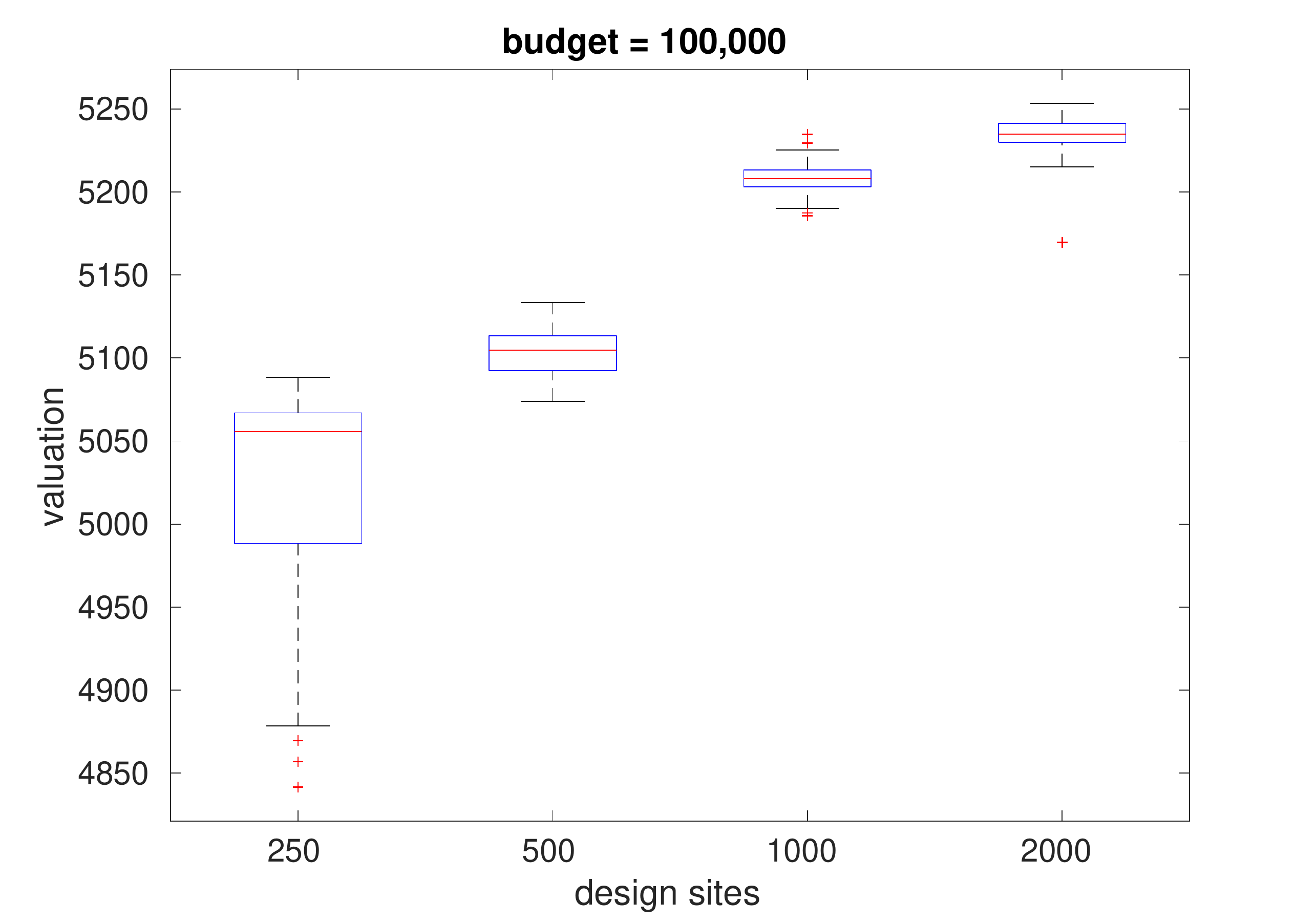}
    \caption{Impact of distinct design sites $N_s$}
    \label{fig:siteEffect}
  \end{subfigure}
\caption{\textit{Left panel:} performance of Mixture GP-2D as a function of $N$. We fix the number of unique design sites $N_s=1000$ and progressively increase the number of replicates $N_b$, reporting the resulting $\hat{V}(0,P,I)$ at $P=6,I=1000$. \emph{Right:} effect of increasing $N_s$ for fixed $N=10^5$. Results are for 20 runs of each algorithm. In each boxplot, the central mark represents the median, the box indicates the 25th and 75th percentiles, and the whiskers represent the most extreme runs.}
\label{fig:designSpecification}
\end{figure}

\textbf{Replicated Designs with GPR. }
Implementation of GPR also requires to manage the tradeoff between the number of design sites $N_s$ and the replication amount $N_b$. The use of replication is necessitated since having more than $N_s > 2000$ distinct sites is  significantly time consuming, at least for the off-shelf-implementation of GPR we used. In the left pane of Figure~\ref{fig:replicationEffect} we consider fixing $N_s$ and varying $N_b$ (hence $N$). While larger simulation budgets obviously improve results, we note that eventually increasing $N_b$ with $N_s$ fixed does not improve the regression quality (although it still reduces standard error).  In the right panel of Figure~\ref{fig:siteEffect} we present the impact of $N_s$ for fixed total budget $N=N_b \cdot N_s$. We find that replication in general sub-optimal and better results are possible when $N_s$ is larger (i.e.~$N_b$ is smaller). To manage the resulting speed/precision trade-off, we recommend taking $N_b \in [20,50]$; for instance when $N = 10^5$ we use $N_s = 2000, N_b =50$ and when $N = 10^4$ we use $N_s = 500, N_b =20$.

\textbf{Take-Aways:} Our experiments suggest the following key observations: (i) Among the inventory-discretized -1D methods, Gaussian process regression outperforms the standard polynomial regression in all cases, and is more robust to ``poor" design or low simulation budget. (ii) Within joint -2D schemes, we continue to observe superior performance of GPR compared to the alternate bivariate regression schemes (LOESS and PR). Moreover, GP-2D is neck-in-neck with  the best-performing GP-1D. We emphasize that efficient implementation of GPR and LOESS relies on batched designs which is another innovation in our DEA implementations. (iii) We also find strong dependence between choice of design and performance. We confirm that best results come from designs, such as the Mixture and Dynamic versions we implemented, that balance filling the input space and targeting the domain where most of $(P_t, \hat{I}_t)$ trajectories lie. Otherwise, plain space-filling or conversely aggressive boundary-following degrade performance. (iv) Between the two parametric methods PR-1D and PR-2D, we find PR-1D to significantly outperform PR-2D irrespective of the design and simulation budget, indicating the advantage of piecewise continuous regression.

\subsection{Gas Storage Modelization with Switching Costs}\label{sec:storage-2}
\label{sec:results_example1_1}

We generalize the previous example by incorporating switching costs. Switching costs make the control map depend on the current regime $m_\ti$ and induce inertia, i.e.~preference to continue with the same regime so as to reduce overall costs. To handle the discrete $m$-dimension, we treat it as $|\mathcal{J}| $ distinct continuation functions, estimated through $|\mathcal{J}|$ distinct regressions. Otherwise, the algorithm proceeds exactly the same way as before. This illustrates the flexibility of DEA to handle a range of problem formulations.

\begin{figure*}[!ht]
        \centering
        \begin{subfigure}[b]{0.3\textwidth}
            \centering
            \includegraphics[width=\textwidth,trim=0.15in 0in 0.15in 0in]{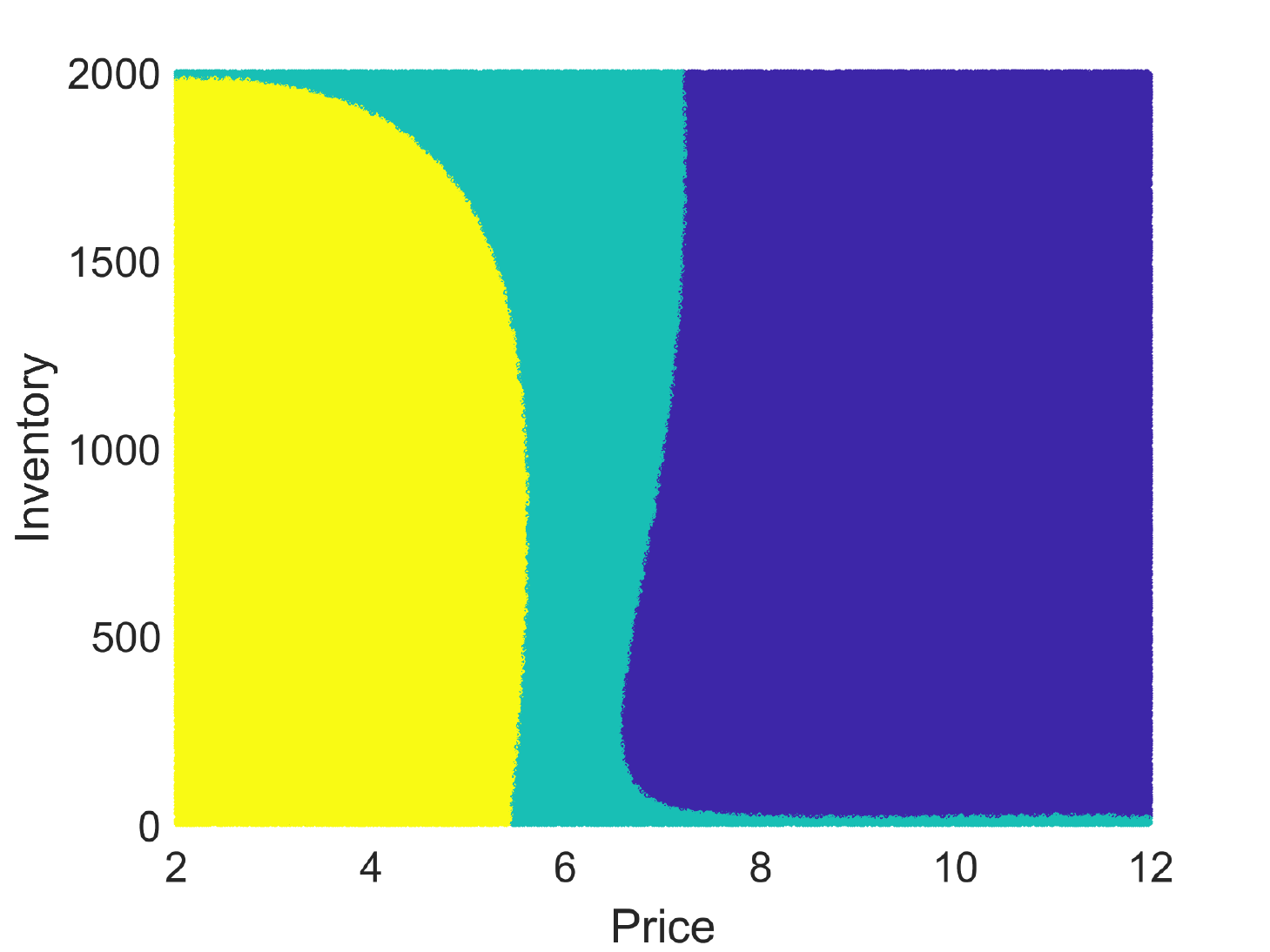}
            \caption[]%
            {{\small $m=+1$ (Inject)}}
        \end{subfigure}
        \hfill
        \begin{subfigure}[b]{0.3\textwidth}
            \centering
            \includegraphics[width=\textwidth,trim=0.15in 0in 0.15in 0in]{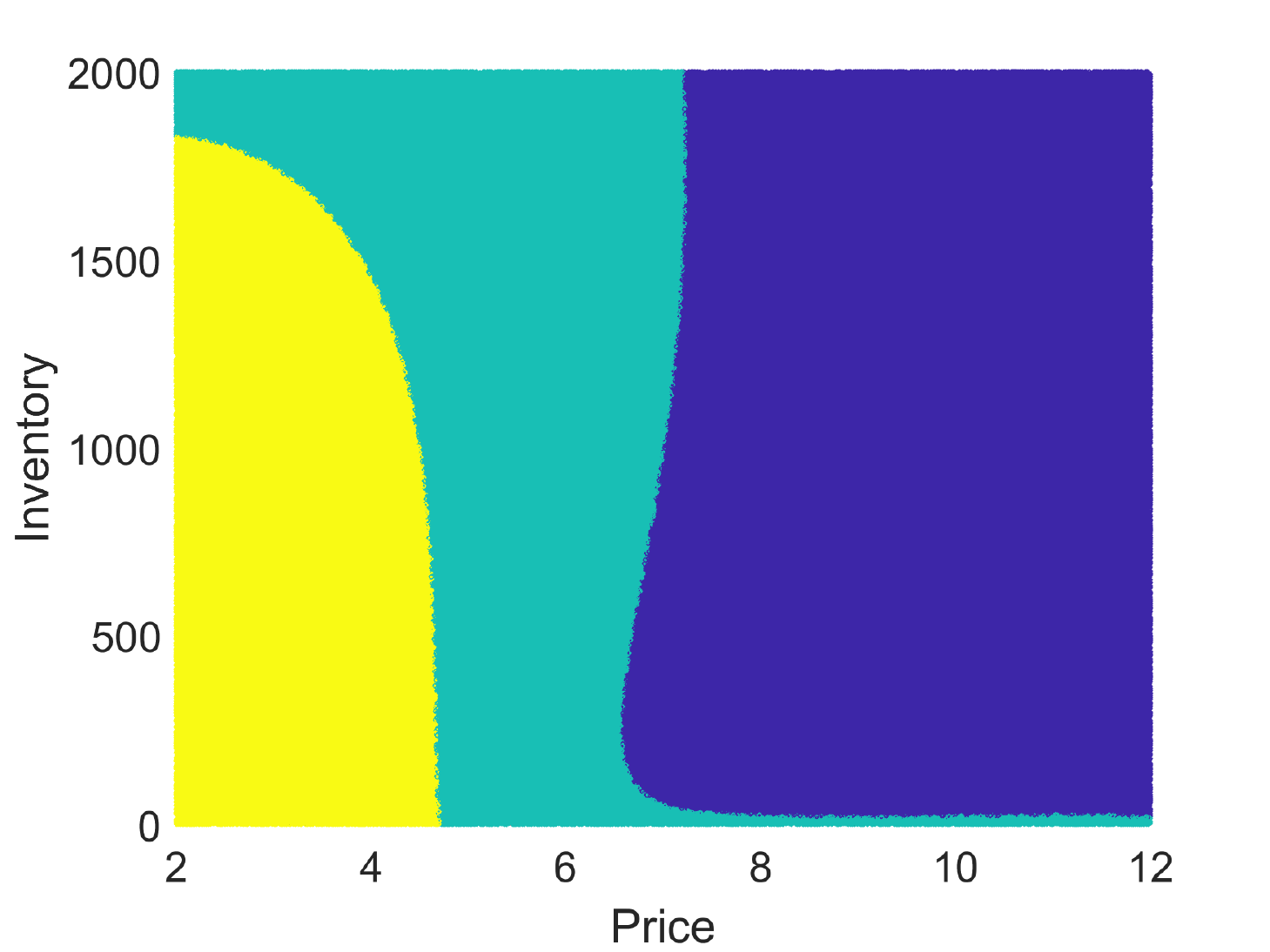}
            \caption[]%
            {{\small $m=0$ (Store)}}
         \end{subfigure}
        \hfill
        \begin{subfigure}[b]{0.3\textwidth}
            \centering
            \includegraphics[width=\textwidth,trim=0.15in 0in 0.15in 0in]{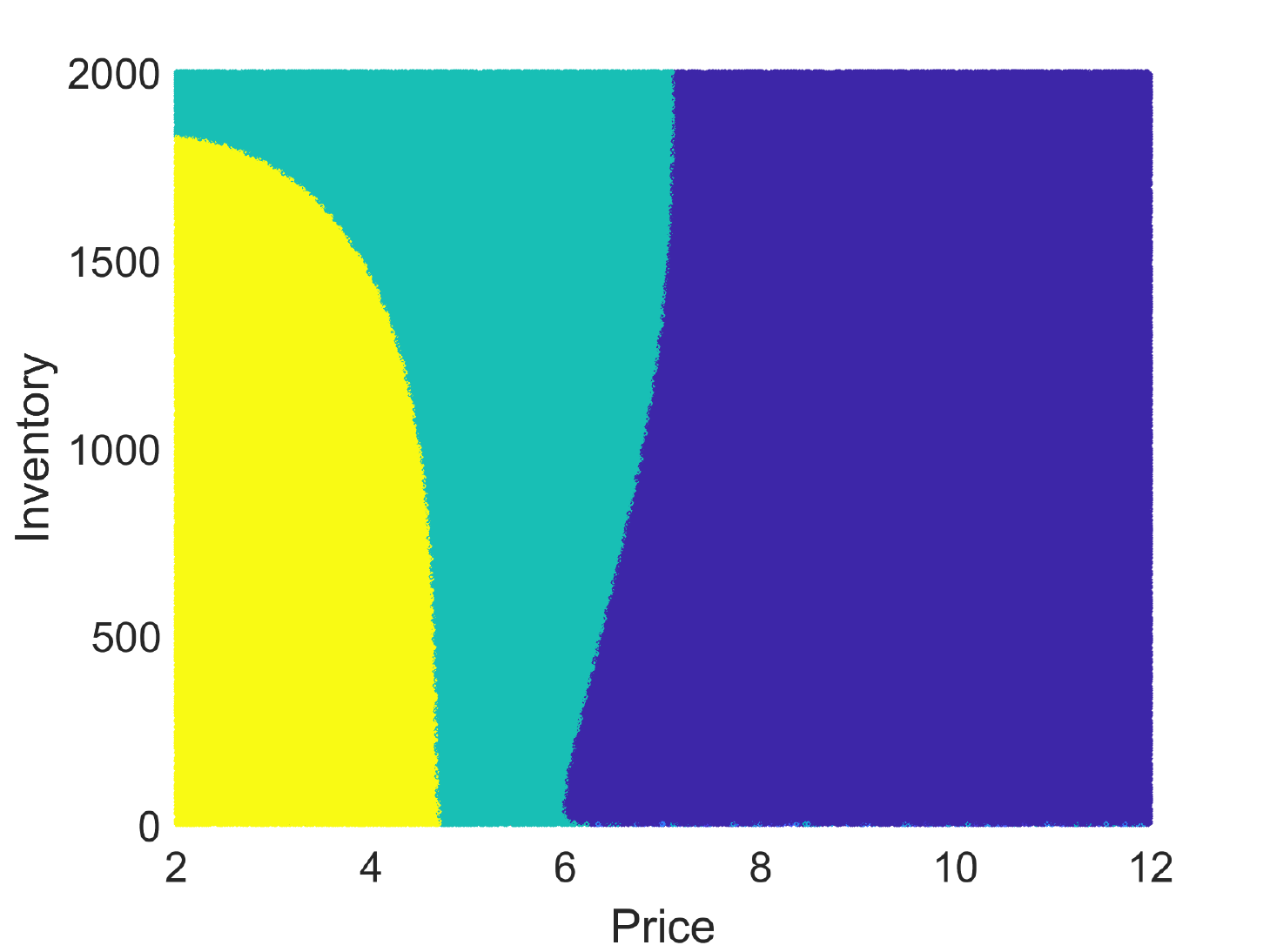}
            \caption[]%
            {{\small $m=-1$ (Withdraw)}}
        \end{subfigure}
            \caption[]%
            {{ The control maps $\hat{m}(t,P,I,m)$ at $t=2.7$ years for the model with switching costs in \eqref{eq:switch-cost}, for $m \in \{ +1,0,-1\}$. The colors are $\hat{m}_{t+\Delta t}=+1$ (inject, light yellow), $\hat{m}_{t+\Delta t} =0$ (store, medium cyan), $\hat{m}_{t+\Delta t} = -1$ (withdraw, dark blue). The solution used GP-2D regression with Mixture design.}}
        \label{fig:controlMapSwitching}
    \end{figure*}

Besides the injection loss through $k_5$, we also add switching cost $K(i,j)$  with the following specification:
\begin{align}\label{eq:switch-cost}
 K(-1,1) = K(0,1) = 15000;  \quad K(1,-1) = K(0,-1) = 5000; \quad\ K(1,0) = K(-1,0) = 0,
\end{align}
i.e.~switching cost depends only on the regime the controller decides to switch to, with switching to injection the costliest and switching to no-action free. In Figure~\ref{fig:controlMapSwitching} we present the policy of the controller for different regimes. The inertia of being in regime $m_\ti=0$ is evident as the corresponding control map has the widest Store region. Effect of $K(i,j)$ is also evident when comparing the Store region of left and center panels. If the controller moves from Inject to Store regime, she finds more resistance while trying to move back to injection due to the switching cost.

In Table \ref{table: DOE2} we present the performance of different design-regression pairs with a $N=40K$ simulation budget. We dropped LOESS from this and the following discussion to simplify the exposition and also because Matlab's implementation of LOESS does not support dimension $d>2$. As expected, introduction of the switching costs leads to lower valuation relative to Table~\ref{table: DOE}. Moreover, the relative behaviour remains similar to the previous section i.e.~space-filling design has the worst performance, Mixture design observes significant improvement, but Dynamic design finally wins the race.

By comparing the valuation in Table \ref{table: DOE2} with Table~\ref{table: DOE} we may infer the impact and number of the switching costs. For example, previously Dynamic GP-1D produced valuation of $\hat{V}(0,6,1000) =\$5,266$ (in thousands), however,  with switching cost it is now $\$5,102K$. The respective loss of $\$ 166K$ can be interpreted as approximately $16$ regime switches on a typical trajectory (assuming $\$10K$ as an average switching cost).

\begin{table}[!ht]
\centering
\begin{tabular}{|l|l|r|}
\hline
Design & Regression & $\hat{V}(0,P_0,I_0)$ ('000s) \\ \hline \hline
\multirow{2}{*}{Conventional} & PR-1D & 4,901 (12) \\
 & PR-2D & 4,654 (11)\\ \hline \hline
\multirow{4}{*}{Space-filling} & PR-1D & 4,663 (16) \\ 
& GP-1D & 4,757 (10) \\ \cline{2-3}
& PR-2D & 4,594 (13) \\ 
 & GP-2D  & 4,879 (22) \\\hline \hline
\multirow{2}{*}{Adaptive-1D} & PR-1D & 4,978 (14) \\ 
& GP-1D & 5,058 (12) \\ \hline
\multirow{2}{*}{Dynamic}  & PR-1D & 4,997 (17) \\
& GP-1D & 5,102 (20) \\  \hline
\multirow{2}{*}{Mixture 2D} & PR-2D & 4,602 (30) \\ 
& GP-2D  & 4,978 (\ 8)   \\ \hline
\end{tabular}
\caption{Valuation of a gas storage facility with switching costs $\hat{V}(0,6,1000)$  of different design-regression pairs with simulation budget of $N=40,000$. Results are averages (standard deviations in brackets) across 10 runs of each algorithm.}
\label{table: DOE2}
\end{table}

\subsection{3D Test Case with Two Facilities}
\label{sec:storage-3d}

An important aspect and motivation for our work has been the question of algorithm scalability in terms of number of input dimensions. In the classical storage problem there are just 2 dimensions: price $P$ and inventory $I$. However, in many contexts there might be multiple stochastic factors (e.g.~the power demand and supply processes in the microgrid example below) or multiple inventories. The respective problem would then be conceptually identical to those considered, except that $\bf{X}$ has $d \ge 3$ dimensions.

Taking up such problems requires the numerical approach to be agnostic to the dimensionality. In terms of existing methods, the piecewise continuous strategy  (such as PR-1D) has been the most successful, but it relies critically on interpolating in the single inventory variable. In contrast, joint polynomial regression is trivially scalable in $d$ but typically performs poorly. Thus, there is a strong need for other joint-$d$ methods that can improve upon PR.

In this section we test how our algorithm and different designs compare as move up to a three-dimensional state variable.
The main idea is to consider a joint model of two gas storage facilities, which leads to three state space variables: price $P_t$, inventory of first storage $I_t^1$ and inventory of second storage $I_t^2$. Each storage facility is independently controlled and operated (so that there are 9 possible regimes $m_t \in \{ -1, 0, 1\} \times \{ -1, 0, 1\}$. Furthermore, the two facilities have identical operating characteristics, each matching those of Section~\ref{sec:storage-1}; it follows that the total value of two such caverns is simply twice the value of a single cavern. 

Overall, the implementation of this setup directly parallels what was done in -2D (highlighting scalability), although we highlight two changes: (i) For the space-filling design we switch to a 3D Sobol sequence; while we do not observe any significant difference in performance between LHS and Sobol sequences for 2D problems, in 3D LHS is less stable (higher variability of $\hat{V}(0,6,1000,1000)$ across runs) and yields estimates that are about \$80K-100K worse than from Sobol designs; (ii) for the mixing weights we take this time 50\%/50\% of sites from space-filling and from empirical distribution i.e.~$\CD = \cP_3(0.5N) \cup \cS_3(0.5N)$. The reason for both modifications is due to lower density of design sites per unit volume compared to previous examples , i.e.~effectively lower budget. Adequate space-filling, best achieved via a QMC design, is needed to explore the relevant input space and fully learn the shape of the 3D continuation function.

\begin{figure*}[!ht]
    \centering
    \includegraphics[height=2.2in,width=5in]{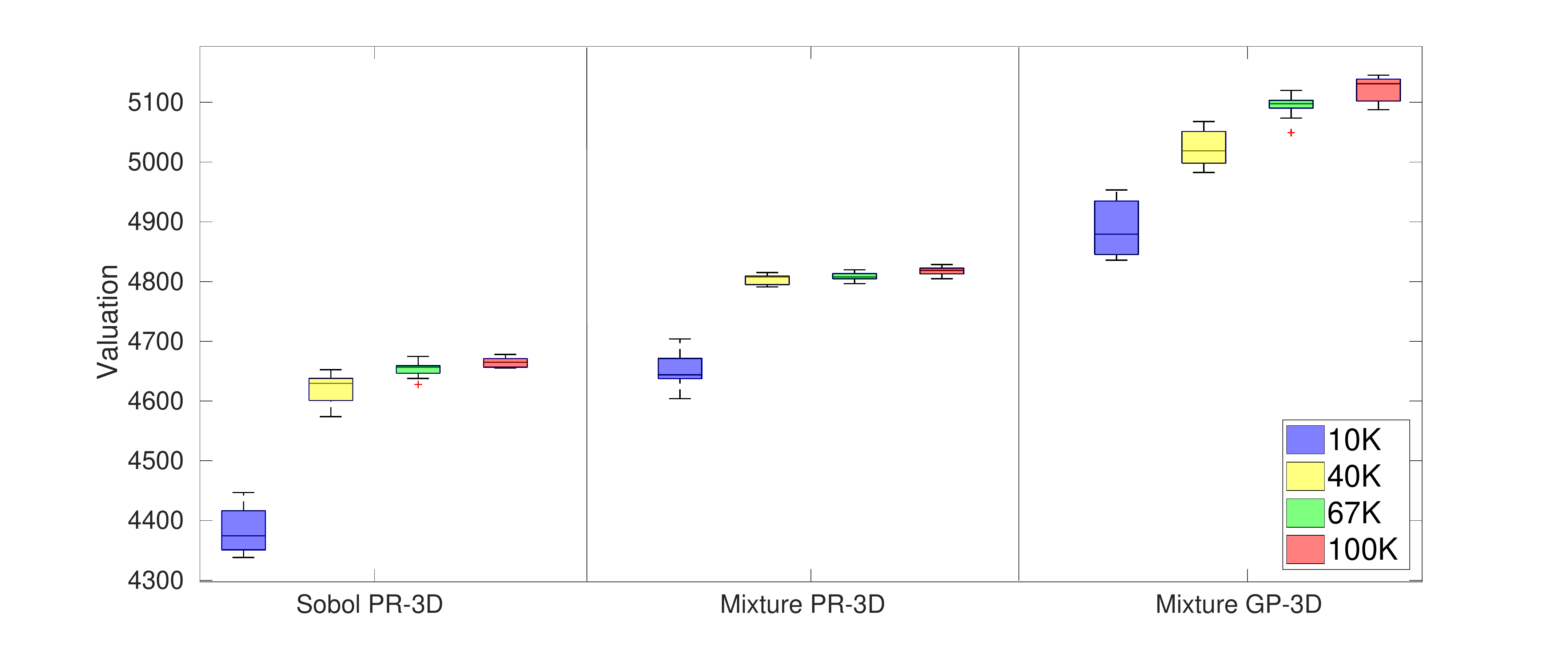}
    \caption{Half of estimated value $\hat{V}(0,6,1000,1000)/2$ for the 3D example with two storage caverns from Section~\ref{sec:storage-3d}. Results are for 10 runs of each algorithm across 4 different simulation budgets $N \in \{10K, 40K, 67K, 100K\}$. Description of the boxplots is same as in Figure~\ref{fig:designSpecification}. }
    \label{fig:storage-3d}
\end{figure*}

In Figure \ref{fig:storage-3d}, we present the performance of PR-3D and GP-3D for Mixture and space-filling designs. 
To ease the comparison, we report half of total value of the two facilities, which should ideally match the original values in Table~\ref{table: DOE}. Not surprisingly, the 3D problem is harder, so for the same simulation budget the reported valuations are lower. For example, at $10^4$ budget, Adaptive GP-3D obtains a valuation $\$251K$ below that of Mixture GP-2D; this gap declines to $\$109K$ as we increase the simulation budget to $N=10^5$.
Moreover, difference between the performance of Mixture and space-filling design is evident even with polynomial regression (Sobol PR vs.~Mixture PR). Mixture design with GP-3D further improves the valuation;  we observe difference of over $\$300K$ comparing Mixture PR and Mixture GP at $N=10^5$ budget.

\section{Microgrid Balancing under Stochastic Net Demand}
\label{sec:microgrid}

\begin{figure*}[!ht]
    \centering
    \includegraphics[height=1.5in, width=0.5\textwidth,trim=0.25in 0.2in 0.25in 0.2in]{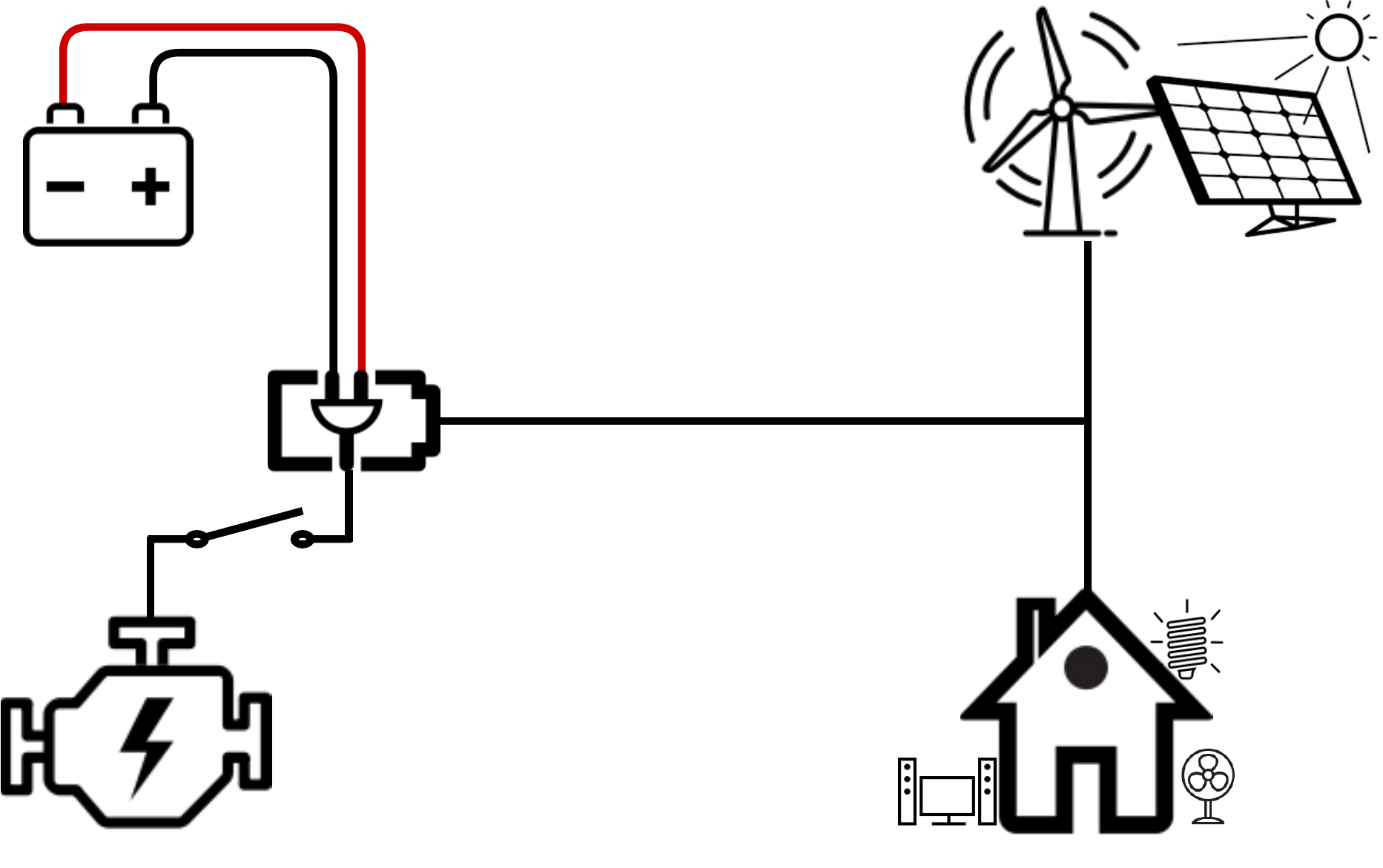}
    \caption{Microgrid topology}
    \label{fig:microgridTopology}
\end{figure*}

In this example, we use the framework of Section~\ref{sec:problem} in the context of a Microgrid, which is a scaled-down version of a power grid comprising of renewable energy sources, a diesel generator, and a battery for energy storage. The microgrid could be isolated or connected to a national grid and the objective is to supply electricity at the lowest cost by efficiently utilizing the diesel generator and the battery to match demand given the  intermittent power production from the renewable sources. The topology of the microgrid considered here is similar to that in \cite{aditya} and presented in Figure~\ref{fig:microgridTopology}.

The exogenous factor corresponds to residual demand $X_\ti = L_\ti - R_\ti$, where $L_\ti$, $R_\ti$ are the demand and output from the renewable source, respectively. We assume  that the microgrid controller bases his policy only on $X$, modeled as a discrete Ornstein-Uhlenbeck process:
 \begin{equation}
X_\tii - X_\ti = \alpha (\underline{X} - X_\ti)\Delta t + \sigma  \Delta W_\ti, \qquad \Delta W_\ti \sim \mathcal{N}(0,\Delta t).
\label{eqn:residualDemand}
\end{equation}

On the supply side, the controller has two resources: a battery (energy storage) and a diesel generator. The state of the battery is denoted by $I_t \in [0, I_{\max}]$ with dynamics given by
\begin{equation}
    I_\tii = I_\ti + a(c_\ti)\Delta t = I_\ti + B_\ti \Delta t ,
    \label{eqn: SOC}
\end{equation}
where $a(c_\ti)$ is interpreted as battery output, driven by $c_\ti$ the diesel output. The latter has two regimes $m_\ti \in \{0,1\}$. In the OFF regime $m_\tii = 0$, $c_\ti(0)=0$. When the diesel is ON, $m_\tii = 1$, its power output is state dependent and given by:
\begin{equation}
  c_\ti(1) = X_\ti \mathbf{1}_{\{ X_\ti>0\} } + B_{\max} \wedge \frac{I_{\max} - I_\ti}{\Delta t},
  \label{eqn:dieselOutput}
\end{equation}
where $B_{\max}>0$ is the maximum power input to the battery.  The power output from the battery is the difference between the residual demand and the diesel output, provided it remains within the physical capacity constraints $[0,I_{\max}]$ of the battery:
\begin{equation}
B_\ti := a(c_\ti) =  -\frac{I_\ti}{\Delta t} \vee\big(B_{\min} \vee (c_{\ti} - X_{\ti}) \wedge B_{\max}\big)\wedge \frac{I_{\max} - I_\ti}{\Delta t}.
\label{eq:battery_output}
\end{equation}

To describe the cost structure, define an imbalance process $S_t = S(c_t,X_t)$:
\begin{equation}
    S_\ti = c_\ti  - X_\ti - B_\ti.
    \label{eqn:imabalance}
\end{equation}
Normally the imbalance is zero, i.e.~the battery absorbs the difference between production and demand. $S_\ti < 0$ implies insufficient supply of power resulting in a \emph{blackout};  $S_\ti > 0$ leads to curtailment or waste of energy. We penalize both scenarios asymmetrically using costs $C_{1,2}$  for curtailment and blackout, taking $C_2 \gg C_1$ in order to target zero blackouts:
\begin{equation}
  \pi(c,X) := - c^{\gamma} - |S| \Big[ C_2\mathbf{1}_{\{S<0\}}+C_1\mathbf{1}_{\{S>0\}} \Big].
  \label{eqn:dieselCost}
\end{equation}
More discussion on the choice of this functional form can be found in \cite{heymann17, aditya}. Furthermore, starting the diesel generator when it is OFF incurs a switching cost $K(0,1)=10$, but no cost is incurred to switch off the diesel generator, $K(1,0)=0$.
The final optimization problem is starting from state $(X_\ti, I_\ti, m_\ti)$ and observing the residual demand process $\textbf{X}_\ti$, to maximize the pathwise value following the policy $\textbf{m}_\ti$, exactly like in \eqref{objFn}.

\subsection{Optimal Microgrid Control} 

The parameters we use are given in Table~\ref{table:microgridParameters}. For the terminal condition we again force the controller to return the microgrid with at least the initial battery charge:
$
W(X_T,I_T) = - 200 \max (I_0 - I_T,0).$
The effect of this penalty is different compared to the gas storage problem in Section~\ref{sec:example1}. Because the controller can only partially control the inventory, we end up with $\hat{I}_T \in [I_0,I_{\max}]$. We use simulation budget of $N=10,000$ and out-of-sample budget of $N'=200,000$. For simplicity of implementation, we use same simulation designs across both $m$-regimes (i.e.~$\CD_k$ is independent of $m$, cf.~Algorithm \ref{algo_Generalized}).

\begin{table}[!ht]
\centering
\begin{tabular}{c} 
\hline
$\alpha = 0.5, \ \underline{X} = 0, \ \sigma = 2$ \\ \hline
$I_{\max} = 10$ (kWh), $B_{\min} = -6, B_{\max} = 6$ (kW), $K(0,1) = 10$, $K(1,0) = 0$ \\ \hline
$C_1 = 5, C_2 = 10^6$, $\gamma = 0.9$, $T=48$ (hours), $\Delta t= 0.25$ (hours) \\ \hline
\end{tabular}
\caption{Parameters for the Microgrid in Section~\ref{sec:microgrid}.}
\label{table:microgridParameters}
\end{table}

\begin{figure}[!ht]
  \centering
  \begin{subfigure}[b]{0.33\textwidth}
    \includegraphics[width=\textwidth,,height=1.8in,trim=0in 0in 0in 0.1in]{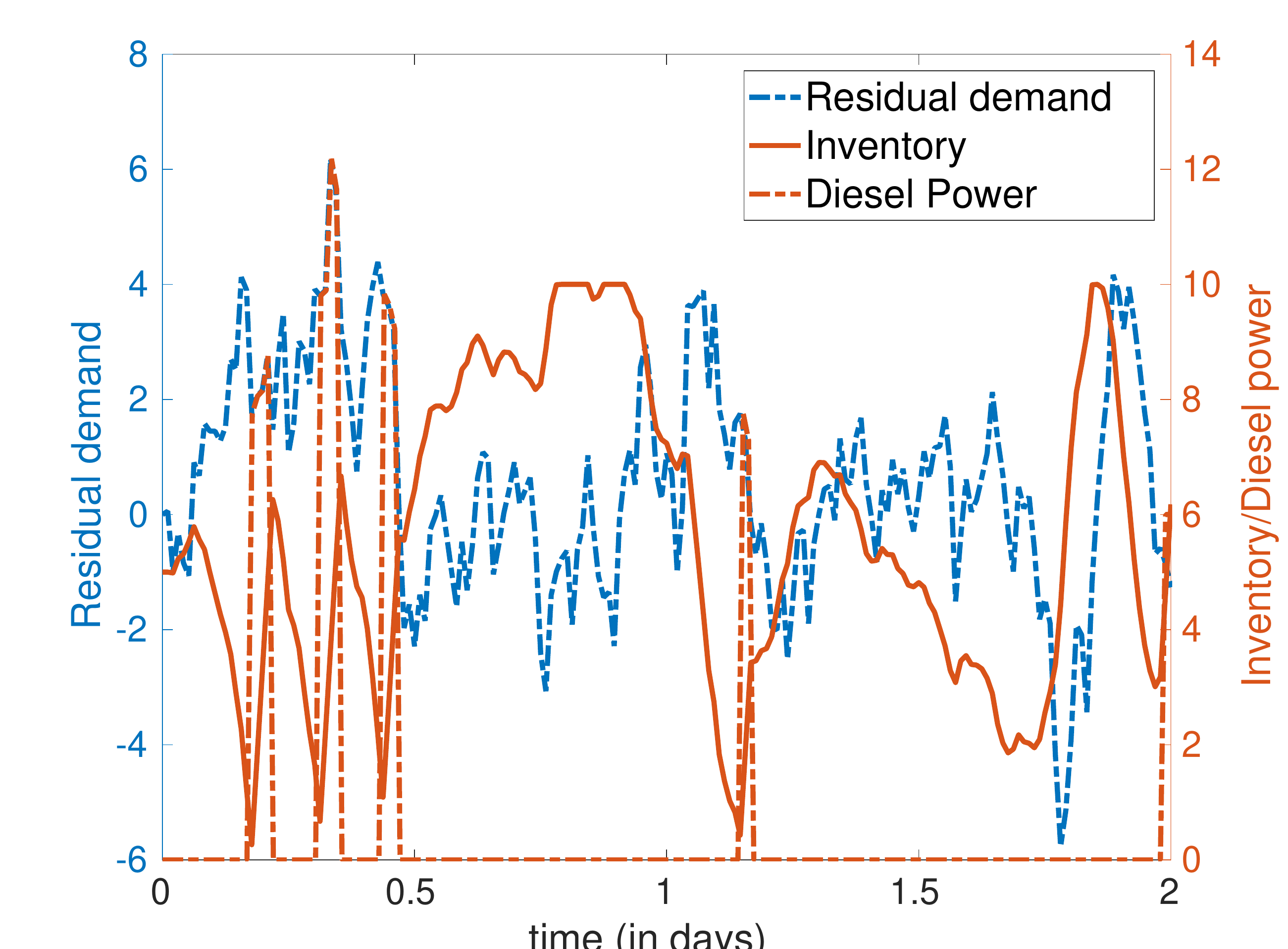}
    \caption{Pathwise trajectory}
    \label{fig:pathwise}
  \end{subfigure}
 \quad
    \begin{subfigure}[b]{0.3\textwidth}
    \includegraphics[width=\textwidth,,height=1.8in,trim=0.1in 0in 0.1in 0.1in]{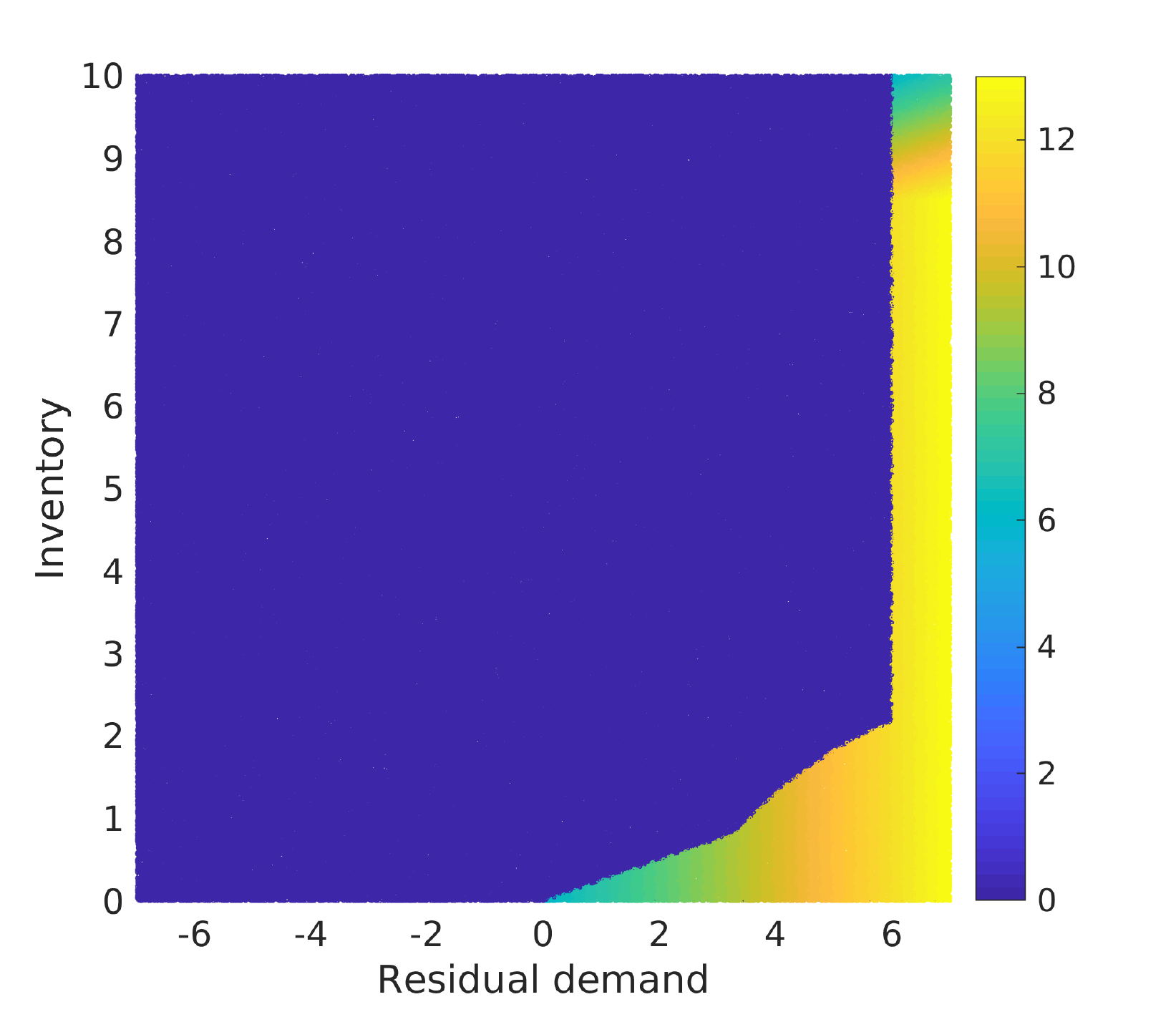}
    \caption{Generator OFF ($m=0$)}
    \label{fig:controlMapOFF}
  \end{subfigure}
 \quad
    \begin{subfigure}[b]{0.3\textwidth}
    \includegraphics[width=\textwidth,height=1.8in,trim=0in 0in 0.1in 0.1in]{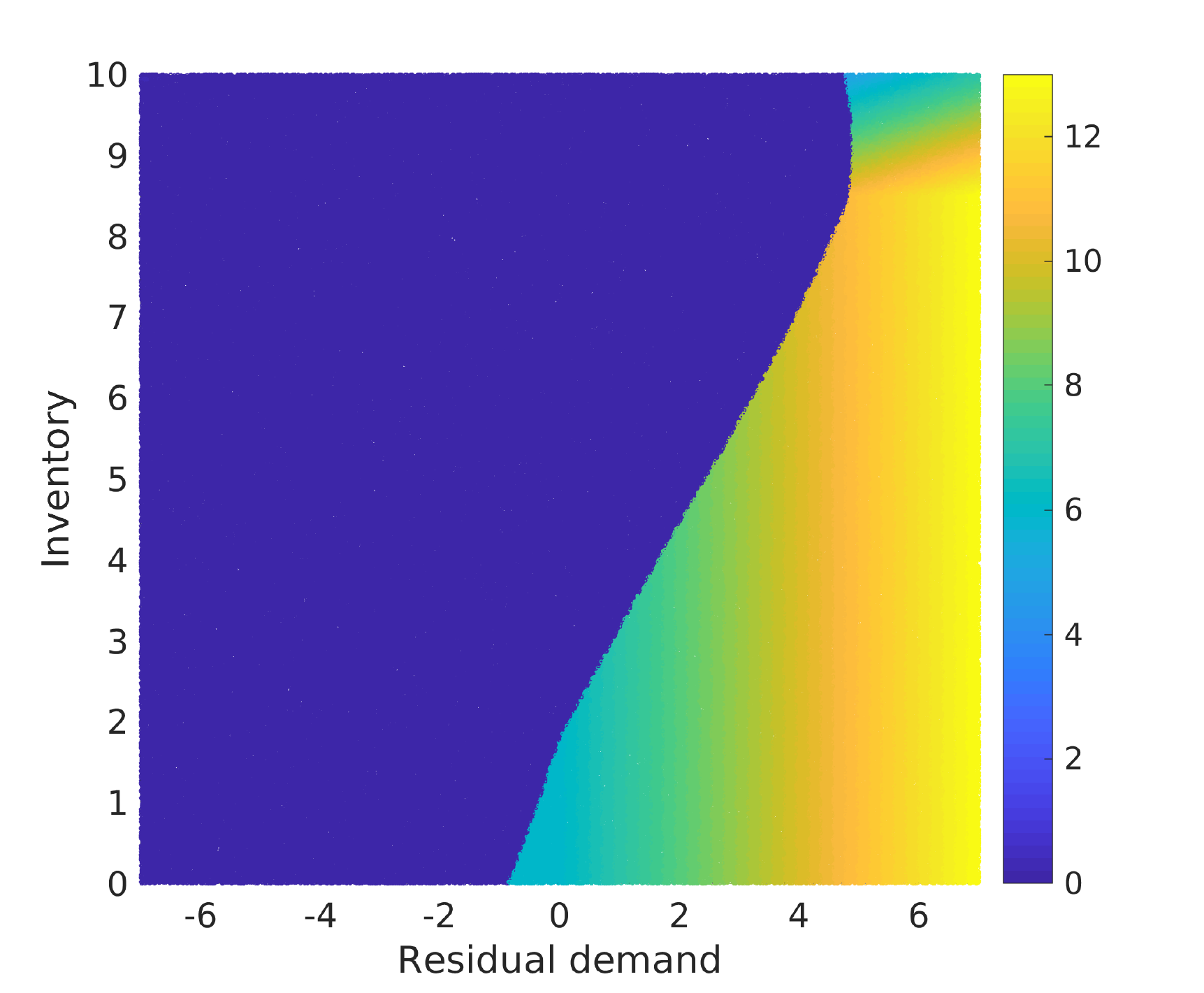}
    \caption{Generator ON ($m=1$)}
    \label{fig:controlMapON}
  \end{subfigure}
\caption{\textit{Left panel:} trajectory of the residual-demand $(X_t)$, corresponding to policy $(c_t)$ and the resultant inventory trajectory $(\hat{I}_t)$. \emph{Middle and right panels:} the control policy $\hat{c}(t,X,I,m)$ at $t=24$ hours. Recall that $c(0)=0$ whenever the diesel is OFF. All panels are based on  GP-2D regression and Mixture design $\CD = \cP_2(0.5N) \cup \cL_2(0.5N)$. }
\label{fig:microgridPathwise}
\end{figure}

Figure \ref{fig:pathwise} illustrates the computed policy $(\hat{m}_t)$ of the microgrid controller  for a given path of residual demand $(X_t)$. The left panel plots the joint trajectory of demand $X_{0:T}$ (left $y$-axis), inventory $\hat{I}_{0:T}$ and diesel output $c_{0:T}$ (both right $y$-axis). The diesel is generally off; the controller starts the diesel generator whenever the residual demand $X_\ti$ is large, or the inventory $I_\ti$ is close to empty. When the generator is on, the battery gets quickly re-charged according to~\eqref{eqn:dieselOutput}; otherwise $\hat{I}$ tends to be decreasing, unless $X_\ti < 0$. The center and right panels of the Figure visualize the resulting policy $c(t,X,I,m)=c(\hat{m}(t,X,I,m))$. Due to the effect of  the switching cost, when the generator is ON (Figure~\ref{fig:controlMapON}), it continues to remain ON within a much larger region of the state space compared to when it is OFF.

\subsection{Numerical Results}

\begin{figure*}[!ht]
    \centering
    \includegraphics[height=2.2in, trim=0in 0.5in 0in 0.5in]{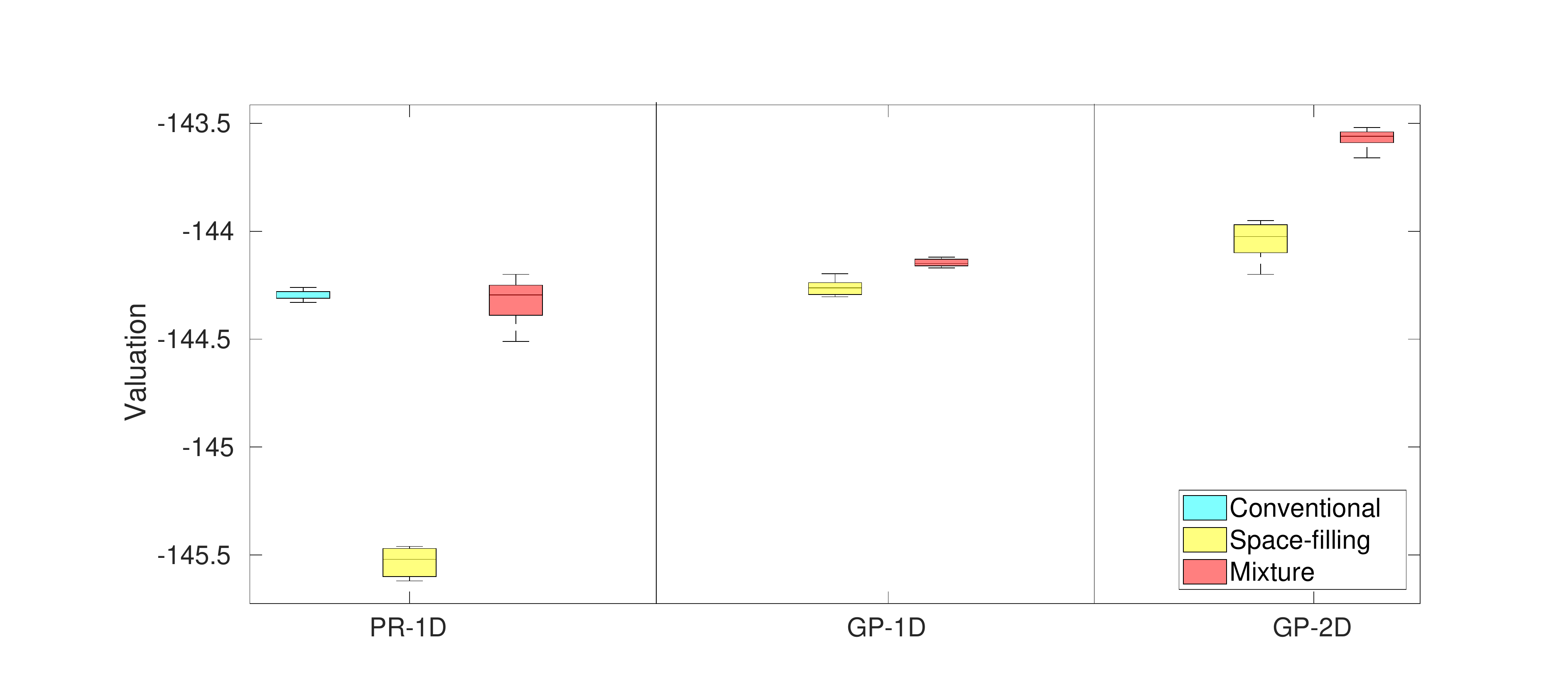}
    \caption{Estimated value $\hat{V}(0,0,5,1)$ for the microgrid example and different design-regression pairs. Results are for 10 runs of each algorithm. For comparison, PR-2D estimated mean valuation was significantly lower at $-153.6$, $-156.4$, $-152.3$ for Conventional, Space-filling and Mixture designs, respectively. Description of the boxplots is the same as in Figure~\ref{fig:designSpecification}.}
    \label{fig:microgridResult}
\end{figure*}

Figure \ref{fig:microgridResult} shows the estimated value  $\hat{V}(0,0,5,1)$  of the microgrid across different designs and regression methods at $N=10,000$. Recall that in this setup, the controller only incurs costs so that $\hat{V} < 0$ and smaller (costs) is better.
The relative performance of the schemes remains similar to Section~\ref{sec:example1}. We continue to observe lower performance of  space filling designs across regression methods. However, GP is more robust to this design change compared to traditional regression methods.
Moreover, GPR dramatically improves upon PR-2D (whose performance is so bad it was left off Figure \ref{fig:microgridResult}). Adaptive design with GP-2D once again produces the highest valuation (lowest cost), and substantially improves upon PR-1D.

\section{Conclusion and Future Work}
\label{sec:conclusion}
The developed DEA template generalizes the existing methodologies used in the sphere of RMC methods for stochastic storage problems. The modularity of DEA allows a wide range of modifications that can enhance current state-of-the-art and improve scalability. In particular, we show several combinations of approximation spaces and simulation designs that are as good or better than any benchmarks (using both -1D and -2D approaches) reported in the literature. Emphasizing the experimental design aspect we show that there is wide latitude in removing memory requirements of traditional RMC by  eliminating the need to simulate global paths. Similarly,  non-parametric regression approaches like GPR, minimize the concern of picking ``correct'' basis functions. Furthermore, we stress the possibility to mix and match different methods. As an example, we illustrated DEA-based valuation of a gas storage facility using different designs, regressions, and budgets across the time-steps.

A natural extension is to consider the setting where the injection/withdrawal rates are continuous. In that case, one must optimize over $c_t$, replacing the $\argmax$ operator with a bonified $\argsup$ over the admissible control set $c \in \cA$. Depending on how switching costs are assessed, one might eliminate the regime $m_t$ entirely, or have a double optimization
$$
V(t,P_t, I_t, m_t) = \max_{m \in \mathcal{J}} \Bigl\{ \sup_{c \in \cA(P_t, I_t, m)} \pi(P_t, c)\Delta t + q(t, P_t, I_{t+\Delta t}(c), m) - K(m_t, m) \Bigr\}.
$$
This could be interpreted as solving a no-bang-bang switching model, which would arise when there is some nonlinear link between profit $\pi$ and $c$, or a nonlinear effect of $c$ on $I_{t+\Delta t}$. Numerically carrying out the inner optimization over $c$ calls for joint regression schemes in order that $\hat{q}(t,\cdot)$ is smooth in $I_{t+\Delta t}$.

A further direction afforded by our template is to move to look-ahead strategies using Remark~\ref{re:look-ahead} for generation of pathwise continuation values. By taking $w > 1$ one may interpolate between the Tsitsiklis-van Roy approach and the Longstaff-Schwartz one, ideally via a data-driven scheme that adaptively selects the look-ahead at each time step. To this end, GP regressions results can be used to quantify the single-step projection errors in $\hat{q}(\ti,\cdot)$.

In a different vein, one may consider modifications where the autonomous/endogenous dichotomy between $(P_t)$ and $(I_t)$ is blurred. For example, a variant of the storage problem arises in the context of hydropower operations, i.e.~controlling a double dammed reservoir that receives inflows from upstream and can release water downstream \cite{delara17,ludkovski10}. Moreover, the reservoir experiences evaporation and/or natural drawdown, irrespective of the operations. In this setup, the inventory $I_t$ experiences stochastic shocks, either due to random inflows (due to precipitation) or random outflows (due to temperature-based evaporation, etc).  Therefore, $I_{t+1}$ is a function of both $I_t, c_t$, and some outside noise (or factor) $O_t$. Moreover, if the dam is large, hydropower management has endogenous stochastic risk, i.e.~the control $c_\ti$ also affects the distribution of the price process $P_\tii$ by modifying the regional supply of energy and hence affecting the supply-demand equilibrium that drives changes in $(P_t)$.

\bibliography{references}
\bibliographystyle{spmpsci}

\clearpage

\appendix

\section{Design specifications}\label{app}

\begin{table}[!ht]
\centering
\begin{tabular}{|l|l|r r r|}
\hline
Design & Regression & Low & Medium & High \\ \hline \hline
\multirow{2}{*}{Conventional} & PR-1D/-2D ($N_P \times N_I$) & $1050\times10$ & $2100\times20$ & $3400\times 30$  \\
& GP-1D/-2D, LOESS ($N_s \times N_b \times N_I$) & $105\times10\times10$ & $210\times 10\times 20$ & $340\times 10\times 30$  \\ \hline \hline
\multirow{4}{*}{Space-filling} & PR-1D ($N_P \times N_I$) & $1050\times 10$ & $2100\times20$ & $3400\times30$   \\
& GP-1D ($N_s \times N_b \times N_I$) & $105\times10\times10$ & $210\times 10\times 20$ & $340\times 10\times 30$  \\
& PR-2D ($N$) & $10500$ & $42000$ & $102000$  \\
 & LOESS/GP-2D ($N_s \times N_b$)  & $500\times 21$ & $1000\times 42$ & $2000\times 51$ \\\hline \hline
\multirow{2}{*}{Adaptive-1D} & PR-1D ($N_P \times N_I$) & $950\times 11$ & $2000\times 21$ & $3300\times 31$ \\ 
& GP-1D ($N_s \times N_b \times N_I$) & $95\times 10\times 11$ & $200\times 10\times 21$ & $330\times 10\times 31$ \\ \hline
\multirow{2}{*}{Mixture-2D}  & PR-2D ($N$) & $10500$ & $42000$ & $102000$  \\
& LOESS/GP-2D ($N_s \times N_b$) & $500\times 21$ & $1000\times 42$ & $2000\times 51$ \\\hline
\multirow{4}{*}{Dynamic}  & GP-1D ($N_{(2)} $) & $150\times 10\times 21$ & $340\times 10\times 31$ & $440\times 10\times 41$ \\
& \hspace{1.2cm}  ($N_{(1)}$ )      & $74\times 10\times 11$ & $168\times 10\times 21$ & $300\times 10\times 31$ \\  \cline{2-5}
& PR-1D + GP-2D ($N_{(2)}$ ) & $2000\times 21$ & $3400\times 31$ & $4400\times 41$  \\
&  \hspace{2.9cm} ($N_{(1)}$ ) & $500\times 21$ & $1000\times 42$ & $2000\times 51$ \\ \hline
\end{tabular}
\caption{Design construction for different methods in Table~\ref{table: DOE} of Section~\ref{sec:storage-1}.  }
\label{table:designSpecification}
\end{table}

\begin{table}[!ht]
\centering
\begin{tabular}{|l|l|}
\hline
$N_I$ & \multicolumn{1}{c|}{Specification} \\ \hline
$11$ & {[}0:100:200, 500:250:1500, 1800:100:2000{]} \\ \hline
$21$ & {[}0:50:200, 400:200:800, 900:50:1100, 1200:200:1600, 1800:50:2000{]} \\ \hline
$31$ & {[}0:25:100, 150, 200:100:900, 950:50:1050, 1100:100:1800, 1850, 1900:25:2000{]} \\ \hline
$\text{else}$ & uniformly spaced \\ \hline
\end{tabular}
\caption{Discretized inventory levels used for Adaptive design for -1D methods in  Table~\ref{table: DOE}.}
\label{table:AdaptiveIlevels}
\end{table}

\begin{table}[!ht]
\centering
\begin{tabular}{|l|l|r r r r|}
\hline
Design & Regression & 10k & 40k & 67k & 100k \\ \hline \hline
\multirow{2}{*}{Sobol/Adaptive} & PR-2D ($N$) & $10500$ & $42000$ & $67500$ & $102000$  \\
 & GP-2D ($N_s\times N_b$)  & $500\times 21$ & $1000\times 42$ & $1500\times 45$ & $2000\times 51$ \\\hline
 \end{tabular}
\caption{Design specifications for different 3D methods of Section \ref{sec:storage-3d}  in Figure \ref{fig:storage-3d}.  }
\label{table:designSpecification3d}
\end{table}

\end{document}